\documentclass[a4paper,11pt, final]{article}
\pdfoutput=1 

\usepackage{jheppub} 

\usepackage[T1]{fontenc} 
\usepackage[utf8]{inputenc}
\usepackage{bm}
\usepackage{siunitx}
\usepackage{slashed}

\newcommand{\ov}[1] {\overline{#1}}
\newcommand{\bp} {\bm{p}}

\newcommand{\brho} {\boldsymbol{\rho}}
\newcommand{\rew} {\mathrm{Re}\, \omega}
\newcommand{\imw} {\mathrm{Im}\, \omega}
\DeclareMathOperator{\Tr}{Tr}
\renewcommand{\Re}{\operatorname{Re}}
\renewcommand{\Im}{\operatorname{Im}}

\title{Parameter space of baryogenesis in the $\nu$MSM}

\author[a]{S. Eijima,}
\author[b]{M. Shaposhnikov,}
\author[b]{I. Timiryasov}

\affiliation[a]{Intituut-Lorentz, Leiden University, Niels Bohrweg 2, 2333 CA Leiden, The Netherlands}
\affiliation[b]{Institute of Physics, Laboratory for Particle Physics and Cosmology,\\
\'{E}cole Polytechnique F\'{e}d\'{e}rale de Lausanne, CH-1015 Lausanne, 
Switzerland}

\emailAdd{Eijima@lorentz.leidenuniv.nl}
\emailAdd{Mikhail.Shaposhnikov@epfl.ch}
\emailAdd{Inar.Timiryasov@epfl.ch}

\abstract{The Standard Model accompanied with two right-handed neutrinos with
masses below the weak scale can explain the observed baryon asymmetry of the Universe.
Moreover, this model is at least partially testable in the forthcoming 
experiments such as NA62, SHiP, and MATHUSLA.
The remarkable progress in understanding of various rates entering the 
kinetic equations describing the asymmetry generation
along with considerable improvements of the numerical procedures
allow us to perform a comprehensive analysis of the parameter space of the model.
We find that the region of parameters leading to successful baryogenesis 
is notably larger than it was previously obtained for light HNLs. 
Our results are presented in a way that they can be readily used
for studies of sensitivity of various experiments searching for the 
right-handed neutrinos 
responsible for the baryon asymmetry of the Universe.
We also present a detailed comparison with the studies by other groups.
}

\begin{document} 
\maketitle
\flushbottom

\section{Introduction} 
\label{sec:introduction}

Neutrino oscillations are among the three experimentally established 
phenomena beyond the Standard Model (SM).
Two others are the baryon asymmetry of the Universe (BAU)
and elusive Dark Matter (DM).

Flavour oscillations of
active neutrinos are prohibited within the canonical SM because
of conservation of individual global lepton numbers. 
The simplest and, probably, the most natural 
way of describing neutrino masses is introduction of 
right-handed neutrinos into the model~\cite{Minkowski:1977sc,GellMann:1980vs,Mohapatra:1979ia,Yanagida:1980xy,Schechter:1980gr,Schechter:1981cv}.
The oscillation data is compatible with the presence of two 
or more right-handed neutrinos. 
In contrast to  the SM particles, there are no symmetries 
prohibiting Majorana mass terms for right-handed neutrinos. The scale of this mass term
is not fixed by neutrino oscillations and can vary by many orders of magnitude.

In refs.~\cite{Asaka:2005an,Asaka:2005pn} it was suggested that 
the minimal extension of the SM with three right-handed neutrinos
with masses \emph{below the electroweak scale}---the $\nu$MSM---can simultaneously
address the problems of neutrino oscillations, dark matter (DM) and BAU.
Two right-handed neutrinos (following the PDG
we will also refer to right-handed neutrinos as heavy neutral leptons or HNLs)
that are responsible for the production of the BAU in the $\nu$MSM
may have masses in the GeV range. They could be searched for
in current and planned experiments. 
The lightest right-handed neutrino may play the role of the 
DM particle~\cite{Dodelson:1993je,Shi:1998km,Dolgov:2000ew,Abazajian:2001nj},
\cite{Asaka:2005an}.

Baryogenesis with GeV scale HNLs suggested in ref.~\cite{Akhmedov:1998qx} 
and refined in ref.~\cite{Asaka:2005pn} has attracted a lot
of attention and a  
significant progress has been achieved recently.
An incomplete list of related works includes~\cite{Shaposhnikov:2006nn,Shaposhnikov:2008pf,Canetti:2010aw,Asaka:2010kk,Anisimov:2010gy,Asaka:2011wq,Besak:2012qm,Canetti:2012vf,Drewes:2012ma,Canetti:2012kh,Shuve:2014zua,Bodeker:2014hqa,Abada:2015rta,Hernandez:2015wna,Ghiglieri:2016xye,Hambye:2016sby,Drewes:2016lqo,Asaka:2016zib,Drewes:2016gmt,Hernandez:2016kel,Drewes:2016jae,Asaka:2017rdj,Eijima:2017anv,Ghiglieri:2017gjz,Eijima:2017cxr,Antusch:2017pkq,Ghiglieri:2017csp}.
The structure of kinetic equations proposed in~\cite{Asaka:2005pn} remained
unchanged, however, understanding of the rates entering into these equations 
has considerably improved 
\cite{Ghiglieri:2016xye,Eijima:2017anv,Ghiglieri:2017gjz}. 
The role of neutrality of electroweak
plasma has been clarified~\cite{Shuve:2014zua,Ghiglieri:2016xye,Eijima:2017cxr}.
Dynamics of the freeze-out of the baryon number has been carefully studied~\cite{Eijima:2017cxr}.

The testability of the model has also drawn a considerable attention from the 
experimental side and the new searches for HNLs were carried out 
\cite{Liventsev:2013zz,Aaij:2014aba,Artamonov:2014urb,Aad:2015xaa,Khachatryan:2015gha,Sirunyan:2018mtv,Boiarska:2019jcw}.
There are several proposals of the experiments which
will be very sensitive to the HNLs of the $\nu$MSM:
NA62 in the beam dump mode~\cite{Drewes:2018gkc}, SHiP~\cite{Alekhin:2015byh} and 
MATHUSLA~\cite{Curtin:2018mvb}.\footnote{Also the recently proposed 
CODEX-b~\cite{Gligorov:2017nwh} and FASER~\cite{Feng:2017uoz,Kling:2018wct} will be probably sensitive
to the HNLs of the $\nu$MSM. Note that indirect searches, such as $\mu \to e \gamma$,
are not sensitive to the $\nu$MSM~\cite{Gorbunov:2014ypa}.
}
Of course, it is important to understand whether the HNLs responsible for baryogenesis
can be found in these experiments.
There were already several studies of the parameter space of the model 
\cite{Canetti:2012vf,Canetti:2012kh,Hernandez:2016kel,Drewes:2016gmt,Drewes:2016jae}
relevant for the current or near-future experiments.

Still, these investigations are not complete. 
In the present work we improve the analysis by
\emph{(i)} using the kinetic equations derived in 
ref.~\cite{Eijima:2017anv}. These equations account for both fermion number conserving 
and violating reactions,
\emph{(ii)} accounting for the neutrality of the electroweak 
plasma and the non-instantaneous freeze-out of the baryon number using methods
suggested in ref.~\cite{Eijima:2017cxr}, and \emph{(iii)} using the fast
numerical code that allows scanning over a wider region of the parameter space. 
We find that the region of parameters leading to the successful baryogenesis 
with light HNLs
is notably larger than it was previously obtained.
The results are presented in a way that they can be used for a detailed study of 
sensitivity of different experiments. 

The paper is organized as follows. First, we introduce the $\nu$MSM and 
the parameters the model in section~\ref{sec:nuMSM}. 
Then we describe the experimentally relevant quantities in 
section~\ref{sec:experimentally_observable_quantities} and
present the cosmologically favourable values of these quantities in section~\ref{sec:results}.
These values are determined by imposing the requirement
of the successful baryogensis. 
In section~\ref{sec:open_access_data_sets}
we  provide all necessary information on the open-access datasets.
All theoretical and technical details
are presented in the subsequent sections.
In section~\ref{sec:production_of_the_baryon_asymmetry} we overview the kinetic
equations derived in ref.~\cite{Eijima:2017anv}. We discuss
our approach for the numerical solution of these equations and 
describe the impact of the improvements in 
section~\ref{sec:numerical_solution_of_the_equations}.
The study of the parameter space is performed in 
section~\ref{sec:scan_of_the_parameter_spaces}. 
Section~\ref{sec:comparison_with_other_works} contains a detailed comparison
with the works~\cite{Canetti:2012vf,Canetti:2012kh,Hernandez:2016kel,Drewes:2016gmt,Drewes:2016jae,Ghiglieri:2017gjz,Ghiglieri:2017csp}.
 We summarise in
section~\ref{sec:conclusions_and_outlook}. 
 Appendix~\ref{sec:mixing_angles_of_hnls_and_active_neutrinos} 
describes the mixings of active neutrinos and HNLs in our parametrization 
of Yukawas.
Appendix~\ref{sec:derivation_of_kinetic_equations} contains the
derivation of the kinetic equations. 
Finally, in appendix~\ref{sec:benchmark_points}
we list several sets of the model parameters along with the corresponding values of 
the BAU. These sets can be used by other groups as benchmarks to compare numerical results.


\section{The \texorpdfstring{$\nu$}{nu}MSM} 
\label{sec:nuMSM}
In this section we fix our notations by introducing 
the Lagrangian of the $\nu$MSM~\cite{Asaka:2005an, Asaka:2005pn}
and the parametrization of Yukawa couplings~\cite{Casas:2001sr,Asaka:2011pb}.
Even though these expressions are well known and have been presented many times,
we list them to make the paper self-consistent.

The Lagrangian of the $\nu$MSM is the usual see-saw 
one~\cite{Minkowski:1977sc,GellMann:1980vs,Mohapatra:1979ia,Yanagida:1980xy,Schechter:1980gr,Schechter:1981cv}
\begin{equation}
	\mathcal{L} = \mathcal{L}_{SM} + i \bar{\nu}_{R_I} \gamma^\mu \partial_\mu 
	\nu_{R_I}
	- F_{\alpha I} \bar{L}_\alpha \tilde{\Phi} \nu_{R_I} - \frac{M_{I J}}{2} 
	\bar{\nu}_{R_I}^c \nu_{R_J} + h.c.,
	\label{Lagr}
\end{equation}
where $\mathcal{L}_{SM}$ is the Lagrangian of the SM, $\nu_{R_I}$ are right-handed neutrinos
labelled with the generation indices $I, J = 1, 2, 3$, $F_{\alpha I}$ is the matrix of 
Yukawa couplings, $L_\alpha$ are the left-handed lepton doublets labelled with
the flavour index $\alpha = e, \mu, \tau$ and $\tilde{\Phi} = i\sigma_2 \Phi^*$,
 $\Phi$ is the Higgs doublet. We work in
a basis where charged lepton Yukawa couplings and the Majorana mass
term for the right-handed neutrinos $M_{I J}$ are diagonal.

In the broken phase, the Higgs field acquires a temperature dependent vacuum expectation value
$\langle \Phi (T) \rangle$, which is $174.1$~GeV at zero temperature.
The Yukawa couplings in the Lagrangian~\eqref{Lagr} lead to the Dirac mass terms
$[M_D]_{\alpha I} = F_{\alpha I} \langle \Phi \rangle$. 
The $6\times 6$ symmetric mass matrix of neutrinos can be diagonalized by a 
complex orthogonal transformation.
We will restrict ourselves to the see-saw limit $|[M_D]_{\alpha I}|\ll M_I$.
In this limit the  active neutrino flavour states are given by
\begin{equation}
\nu_{L_\alpha} = U^{PMNS}_{\alpha i} \nu_i + \Theta_{\alpha I} N_I^c,
	\label{mixing}
\end{equation}
where $U^{PMNS}$ is the Pontecorvo-Maki-Nakagawa-Sakata (PMNS) matrix~\cite{Pontecorvo:1957qd,Maki:1962mu}, $\nu_i$ are mass eigenstates of active neutrinos, 
$N_I$ are the mass eigenstate
of HNLs. The active-sterile mixing matrix 
in the leading order of the see-saw mechanism is
\begin{equation}
    \Theta_{\alpha I} = \frac{\langle \Phi \rangle F_{\alpha I}}{M_I}.
    \label{Theta}
\end{equation}

The parameters of the theory~\eqref{Lagr} are restricted by the see-saw mechanism
since one has to reproduce the observed values of the mass differences and mixing angles for the active neutrinos~\cite{Esteban:2016qun}.\footnote{The NuFIT
group has recently released an updated global analysis of neutrino oscillation measurements, \emph{NuFIT 3.2 (2018), www.nu-fit.org}. In our analysis,
we use these updated data.
The most important update of v3.2 is the $3 \sigma$ bound 
on the value of the Dirac phase $\delta$. In the inverted hierarchy case
$\delta$ is no longer compatible with zero. We will comment on this
in section~\ref{sec:scan_of_the_parameter_spaces}.}
A convenient parametrization of the Yukawa couplings which automatically
accounts for these observables was proposed by Casas and Ibarra
 in ref.~\cite{Casas:2001sr}. The application of the Casas-Ibarra parametrization
to the $\nu$MSM has been studied in ref.~\cite{Asaka:2011pb}.
In the matrix form, the
Yukawa couplings entering the Lagrangian~\eqref{Lagr} read 
(in the notations of refs.~\cite{Eijima:2017anv,Eijima:2017cxr})
\begin{equation}
    F = \frac{i}{\langle \Phi (0) \rangle} U^{PMNS} m_\nu^{1/2} \Omega m_N^{1/2},
    \label{CasasIbarra}
\end{equation}
where $m_\nu$ and $m_N$ are the diagonal mass 
matrices of the three active neutrinos and HNLs correspondingly.
The matrix $\Omega$ is an arbitrary complex orthogonal $\mathcal{N}_\nu \times \mathcal{N}_N$ matrix,
where $\mathcal{N}_\nu$ is the number of left-handed neutrinos and $\mathcal{N}_N$ is the number
of right-handed neutrinos.

In the $\nu$MSM, the lightest HNL $N_1$ is the dark matter particle.
A combination of Lyman-$\alpha$ and X-ray constraints puts strong bounds on the 
magnitude of its Yukawa couplings, see~\cite{Boyarsky:2009ix} and references therein. 
As a result, $N_1$
is almost decoupled and does not contribute to the see-saw masses of active neutrinos.
Therefore, the masses and mixings of active neutrinos correspond to the case
of two HNLs. In this case the matrix $\Omega$ can be chosen in the form
\begin{align}
\Omega &=   \left(
\begin{array}{cc}
 0 & 0 \\
 \cos \omega &  \sin \omega \\
 -\xi  \sin \omega & \xi  \cos \omega \\
\end{array}
\right)\quad\text{for NH},\\
\Omega &=   \left(
\begin{array}{cc}
 \cos \omega &  \sin \omega \\
 -\xi  \sin \omega & \xi  \cos \omega \\
 0 & 0 \\
\end{array}
\right)\quad\text{for IH},
\end{align}
with a complex mixing angle $\omega$. The sign parameter is $\xi= \pm1$. We fix it to
be $\xi = +1$, 
since the change of the sign of $\xi$ can
be compensated by $\omega \to - \omega$ 
along with $ N_3 \to - N_3$~\cite{Abada:2006ea}.
Throughout this work, we will use the abbreviations NH and IH to refer to
the normal and inverted hierarchy of neutrino masses.
In what follows it is  convenient to introduce
\begin{equation}
    X_\omega=\exp( \imw).
\end{equation}

In the case of two right-handed neutrinos, the PMNS matrix contains only two $CP$-violating phases, 
one Dirac $\delta$ and one Majorana $\eta$, see Appendix 
\ref{sec:mixing_angles_of_hnls_and_active_neutrinos}
for the details of our parametrization of the PMNS matrix.
Two Majorana masses in~\eqref{Lagr} can be parametrized by the common mass $M$
and the Majorana mass difference $\Delta M$. Note that the \emph{physical} mass
difference controlling the oscillations of two HNLs
is a sum of $\Delta M$ and a term proportional to the product of Yukawa couplings
with $\langle \Phi \rangle$. The expression for this mass difference can be found 
in ref.~\cite{Shaposhnikov:2008pf}.

We end up with six free parameters of the theory
which are listed in table~\ref{table_parameters} along
with their ranges considered in this work.
\begin{table}[htb!]
\begin{center}
  \begin{tabular}{| c | c | c | c | c | c |}
    \hline
   $M$, GeV & $\log_{10} (\Delta M/\mbox{GeV})$  & $\imw$ & $\rew$ & $\delta$ &
    $\eta$\\ \hline
   $[0.1 - 10]$  & $[-17,-1]$  & $[-7,7]$ & $[0, 2\pi]$ &$[0, 2\pi]$ & $[0, 2\pi]$\\ 
    \hline
  \end{tabular}
\end{center}
\caption{\label{table_parameters} Parameters of the theory:
common mass; mass difference; $\imw$; $\rew$;
Dirac and Majorana phases.
In the second line we indicate the ranges of these parameters which were considered in this work.}
\end{table}
The common mass $M$ of HNLs is restricted to the $[0.1 - 10]$~GeV interval.
The smaller masses are in tension with the Big Bang nucleosynthesis 
(BBN)~\cite{Ruchayskiy:2012si}. Heavier HNLs, which we do not consider in this
work, deserve a separate study.
The ranges of the Majorana mass difference $\Delta M$ and $\imw$ are
determined by the a posteriori requirement of generating enough BAU.
The real part of the complex angle $\omega$ plays a role of a phase, therefore
it is enough to restrict it, along with the single Majorana
phase, to the interval $[0, 2 \pi]$. The range of the Dirac phase $\delta$
is somewhat more subtle since it was restricted in the recent
global analysis of neutrino oscillation measurements. We will 
comment on this in section~\ref{sec:scan_of_the_parameter_spaces}.

Note that the relation~\eqref{CasasIbarra} is not an isomorphism, i.e. more than
one set of the parameters lead to the same Yukawas $F_{\alpha I}$
and therefore are physically equivalent.
Still, the parametrization~\eqref{CasasIbarra} 
spans all possible values of Yukawas compatible with the oscillation data.
Dependence of the resulting BAU on the Yukawas (and parameters in table~\ref{table_parameters})
is in general very complicated. 
Therefore a thorough study of the parameter space of the
model is required in order to determine the boundary of the region where successful
baryogenesis is possible in terms of the experimentally 
interesting quantities.

\section{Experimentally observable quantities} 
\label{sec:experimentally_observable_quantities}
The parameters listed in table~\ref{table_parameters} are useful for the theoretical understanding of the model, but the last four of them cannot be directly measured.
In this section, we discuss the experimentally observable quantities and
their relations to the parameters of the model.

\subsection{The total mixing} 
\label{sub:the_total_mixing}

The formula~\eqref{mixing} establishes the basis for experimental searches
of the HNLs. It shows that an amplitude
of a process involving HNL $N_I$ is equal to the analogous
amplitude involving active neutrino $\nu_\alpha$  multiplied 
by $\Theta_{\alpha I}$.

In order to understand, how weakly HNLs are coupled to the SM in general,
it is helpful to sum $|\Theta_{\alpha I}|^2$ over flavours of active neutrinos
and over $I=2,3$. This defines the total mixing
\begin{equation}
    |U|^2 \equiv \sum_{\alpha I} |\Theta_{\alpha I}|^2 =
    \frac{1}{2M}\left[(m_2 + m_3)\left(X_\omega^2 + X_\omega^{-2}\right)
    + \mathcal{O}\left(\frac{\Delta M}{M} \right)\right],
    \label{U2}
\end{equation}
where $m_{2,3}$ are masses of active neutrinos and the normal hierarchy
(NH) of the active neutrino masses is assumed (the inverted hierarchy
(IH) case can be obtained by replacing $m_2\to m_1, m_3\to m_2$
in~\eqref{U2}).
The total mixing~\eqref{U2} controls the amount
of HNLs produced in an experiment and the lifetime of these HNLs.

\subsection{Individual mixings} 
\label{sub:individual_mixings}

The total mixing~\eqref{U2} is useful to quantify interactions
of the HNLs with the SM particles, however, it is not sufficient for
determining sensitivity of experiments to the HNLs.
Therefore we also consider individual, or flavoured, mixings.

To clarify the role of flavoured mixings, let us consider, e.g. the SHiP
experiment~\cite{Anelli:2015pba,Alekhin:2015byh}.
This is the beam damp-type experiment. An intense proton beam from the SPS accelerator
hits the target. The main detector consists of a large empty decay volume
with calorimeters and trackers at the end.
In the SHiP set-up, HNLs are supposed to be produced mostly in decays of heavy mesons
and the observational signatures consist of boosted charged particles originating from a vertex in the empty volume.
The production is proportional to the partial decay width of a heavy meson into an HNL
$\Gamma(H \to N_I \ell_\alpha)$, which is in turn proportional to the 
$|\Theta_{I \alpha}|^2$.
It is important to note that the HNL production channels with different accompanying leptons $\ell_\alpha$ are in principle distinguishable.
In the SHiP, they could be discriminated if the mass of HNLs is
close to upper bounds of kinematically allowed regions.
Let us illustrate this in an example. Suppose that one observes a
decay of an HNL with the mass exceeding $m_{B_c}-m_\mu$ to a muon.
This means that the HNL was produced along with an electron in the process
$B_c\to N_I e$ since the process $B_c\to N_I \mu$ is kinematically forbidden.

The decay widths of HNLs are in turn proportional to $|\Theta_{J \beta}|^2$. The channels with charged leptons $\ell_\beta$
in the final state are also distinguishable. 
In the example above the product of $|\Theta_{I \alpha}|^2$ and
$|\Theta_{I \beta}|^2$ is important.

Therefore, individual mixings are  phenomenologically relevant.
Notice also that for the mass difference range which we are studying here the characteristic oscillation length
is several orders of magnitude smaller than the length SHiP shielding and fiducial volume.\footnote{The case when the oscillations
of HNLs are important will be addressed in a separate study.}
Namely, the oscillation length $100$ m coincides to $\sim 10^{-7}$~eV 
\emph{physical} mass difference.
 Therefore
for the lepton $\ell_\alpha$ in the target and $\ell_\beta$ in the detector one has to sum incoherently over $I, J$.
So the total dependence on mixings is 
$$
\sum_{I, J = 2,3}|\Theta_{I \alpha}|^2\cdot |\Theta_{J \beta}|^2 =|U_\alpha|^2\cdot|U_\beta|^2.
$$

The individual mixings as functions of the parameters are given by
\begin{equation}
	|U_\alpha|^2 \equiv \sum_{I= 2,3}|\Theta_{I \alpha}|^2 =
	\frac{1}{2M} \left( |C_\alpha^+|^2X_\omega^2  + |C_\alpha^-|^2X_\omega^{-2}
	+ \mathcal{O}\left( \frac{\Delta M}{M} \right)  \right),
	\label{Ua2}
\end{equation}
where the combinations $C_\alpha^\pm$ for the NH case are
\begin{equation}
	C_\alpha^\pm = i U^{PMNS}_{\alpha 2}\sqrt{m_2} \pm 
	\xi U^{PMNS}_{\alpha 3}\sqrt{m_3}
	\label{Cpm_nh}
\end{equation}
and for the IH case
\begin{equation}
	C_\alpha^\pm = i U^{PMNS}_{\alpha 1}\sqrt{m_1} \pm 
	\xi U^{PMNS}_{\alpha 2}\sqrt{m_2},
	\label{Cpm_ih}
\end{equation}
where  $U^{PMNS}_{\alpha i}$ with $i=1,2,3$ are the elements of the PMNS 
matrix (should not be confused with  $|U_\alpha|^2$).



\section{Cosmologically motivated values of mixings} 
\label{sec:results}
In this section, we present our main results, namely,
the values of the total and individual mixings of HNLs with active neutrinos
for which the observed BAU can be explained by the $\nu$MSM.
We describe how these results were obtained in a separate section~\ref{sec:scan_of_the_parameter_spaces}.

The value of the BAU can be characterised in different ways.
Throughout this work, we use the variable $Y_B = n_B/s$, where
$n_B$ is the baryon number density 
(particles minus antiparticles) and $s$ is the entropy density.
The observed value is 
$Y_B^{obs} = (8.81 \pm 0.28) \cdot 10^{-11}$~\cite{Patrignani:2016xqp}.
For each set of the model parameters, we  numerically find the value of $Y_B$.
We are interested in the regions of the parameter space where one 
can reproduce the observed value $Y_B^{obs}$.

\begin{figure}[htb!]
\centerline{
    \includegraphics[width = 0.5\textwidth]{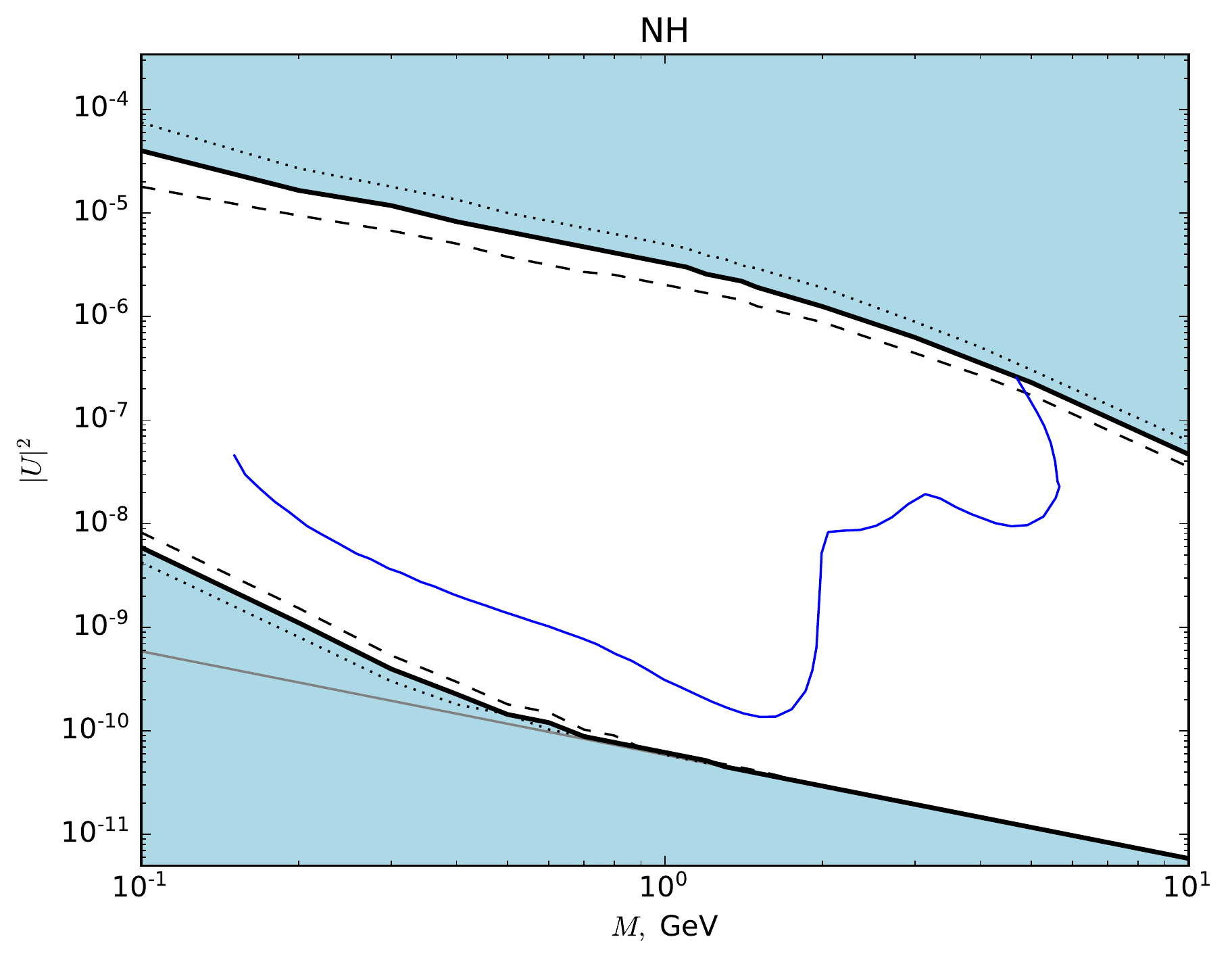}
    \includegraphics[width = 0.5\textwidth]{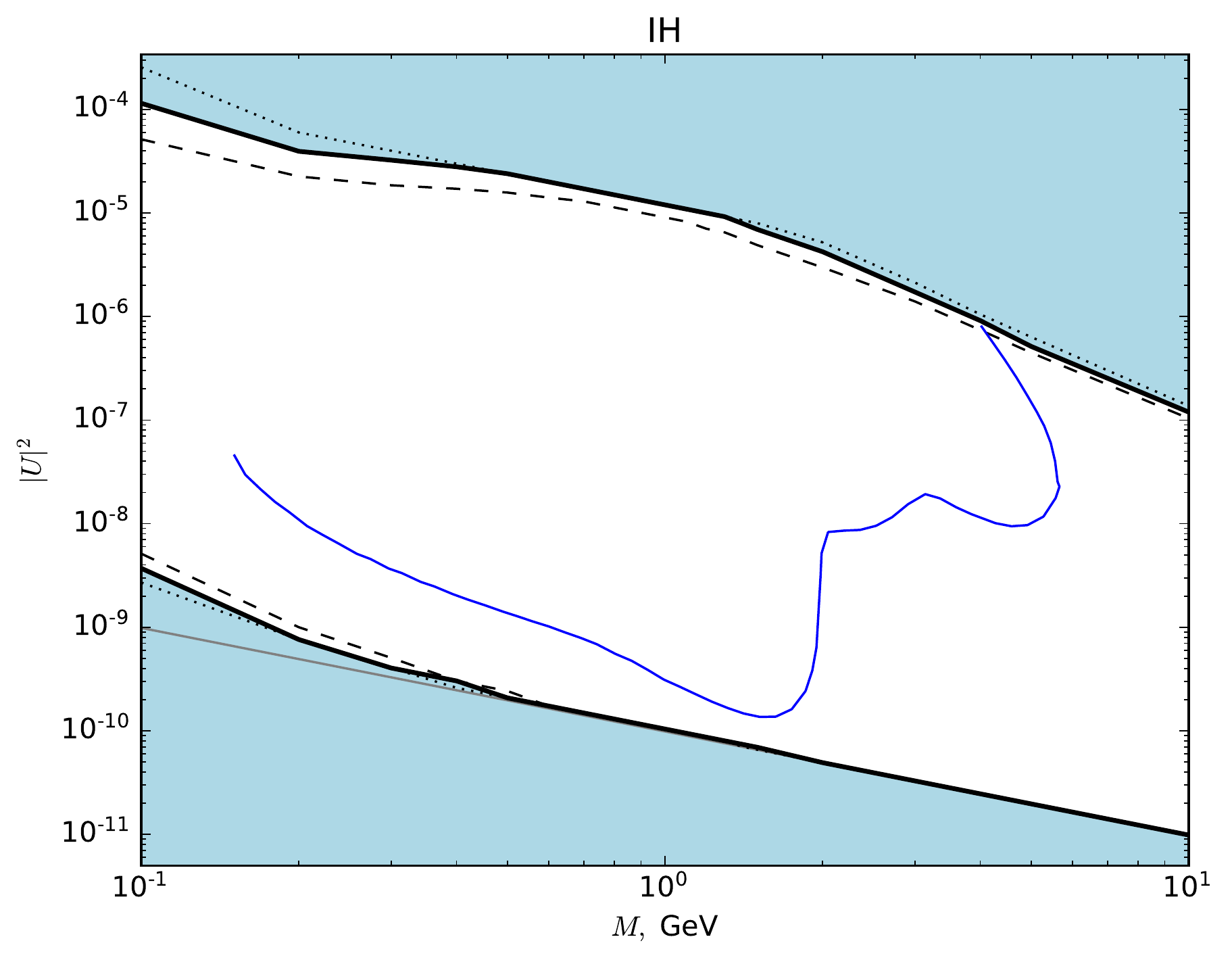}
}
\caption{\label{U2_bounds}
Within the white regions it is possible to reproduced the observed
value of the BAU (black solid curves).
The dashed and dotted curves demonstrate how large the possible 
theoretical uncertainties could be.
Namely, the dashed curves correspond to the  condition
$Y_B \ge 2\cdot Y_B^{obs}$, whereas the dotted lines correspond
to $Y_B \ge Y_B^{obs}/2$ accounting for the factor of $2$ uncertainty in the
computation of the BAU. The thin grey lines show the see-saw limit, i.e.
it is impossible to obtain the correct masses of active neutrinos
below these lines.
The blue line shows the projected sensitivity of the SHiP experiment 
ref.~\cite{ship_new} as
presented in ref.~\cite{Bondarenko:2018ptm}.
 \emph{Left panel}: normal hierarchy, \emph{right panel}: inverted hierarchy.}
\end{figure}

The regions of successful baryogenesis are shown in figure~\ref{U2_bounds}. 
In order to indicate
how large can be the effect of the theoretical 
uncertainties in BAU computation, discussed in 
section~\ref{sec:production_of_the_baryon_asymmetry},
we show the borders of the regions where one can generate
$2\cdot Y_B^{obs}$, and $Y_B^{obs}/2$.

The cosmologically favoured region of the parameter space is  larger for light HNLs
 than
it was previously recognized (see also the discussion in 
section~\ref{sec:comparison_with_other_works}, in particular,
figure~\ref{fig:comparison}). The fact that successful baryogenesis
is possible for quite large values of the mixings rises the question about
the upper bounds of
sensitivity of the direct detection experiments.
To illustrate this point, we  estimate the lifetime of an HNL using
expressions for the decay rates of HNLs from 
ref.~\cite{Gorbunov:2007ak}\footnote{Note that this work was  updated
recently~\cite{Bondarenko:2018ptm} with new channels added, so our estimate is
conservative.}. For instance, let us consider an HNL with the mass
$M=5$~GeV and mixings close to the upper boundary in figure~\ref{U2_bounds}.
For such an HNL the lifetime is of the order of $5\cdot 10^{-9}$~s. 
Estimating the gamma factor 
to be $\simeq 10$ we see that the decay length in the lab frame
is less than $15$~m.
This implies that, e.g., in the SHiP experiment this HNL will decay well
before the detector. 
Therefore it might be interesting to revisit the current experimental bounds
on HNLs. 

Let us note in passing that there also exist bounds from the Big Bang 
nucleosynthesis~\cite{Gorbunov:2007ak,Boyarsky:2009ix,Canetti:2012kh}. 
The question of the derivation of such bounds
has been addressed in details for HNLs with the mass below $140$~MeV
in ref.~\cite{Ruchayskiy:2012si}. For heavier Majorana HNLs an accurate derivation
is still missing. 

Results for the individual mixings $|U_\alpha|^2$
and products $|U_\alpha|\cdot|U_\beta|$ are presented in figures~\ref{UaUb_NH} and~\ref{UaUb_IH}.
{
\begin{figure}[htb!]
\includegraphics[width = 1.0\textwidth]{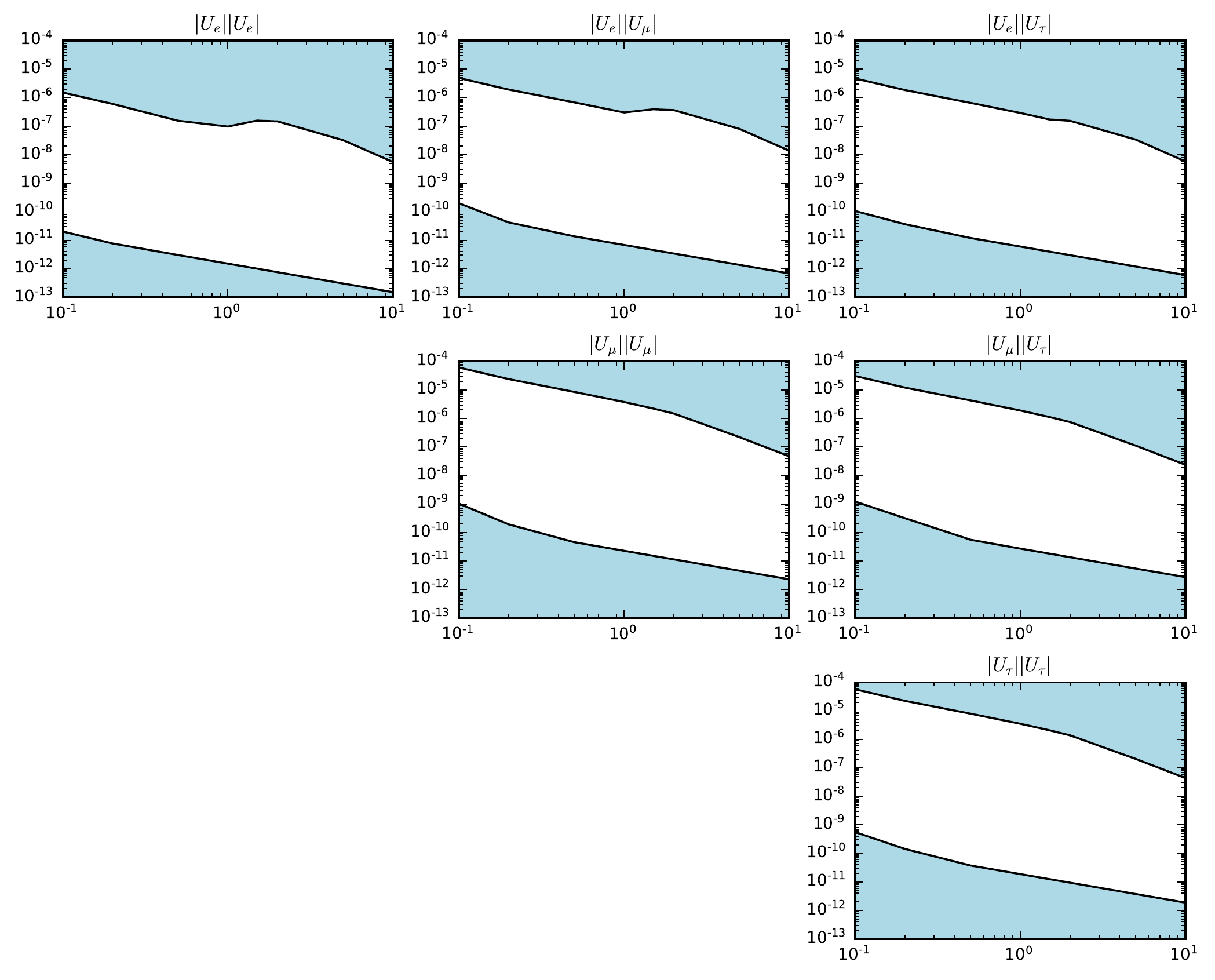}
\caption{\label{UaUb_NH} Cosmologically motivated regions of the individual mixings $|U_\alpha|^2$
and products $|U_\alpha|\cdot|U_\beta|$. Within the white regions it is possible to reproduced the observed
value of the BAU. The common mass is shown in the horizontal
axes, whereas the vertical axes show the corresponding product of mixings.
NH case.}
\end{figure}
\begin{figure}[htb!]
\includegraphics[width = 1.0\textwidth]{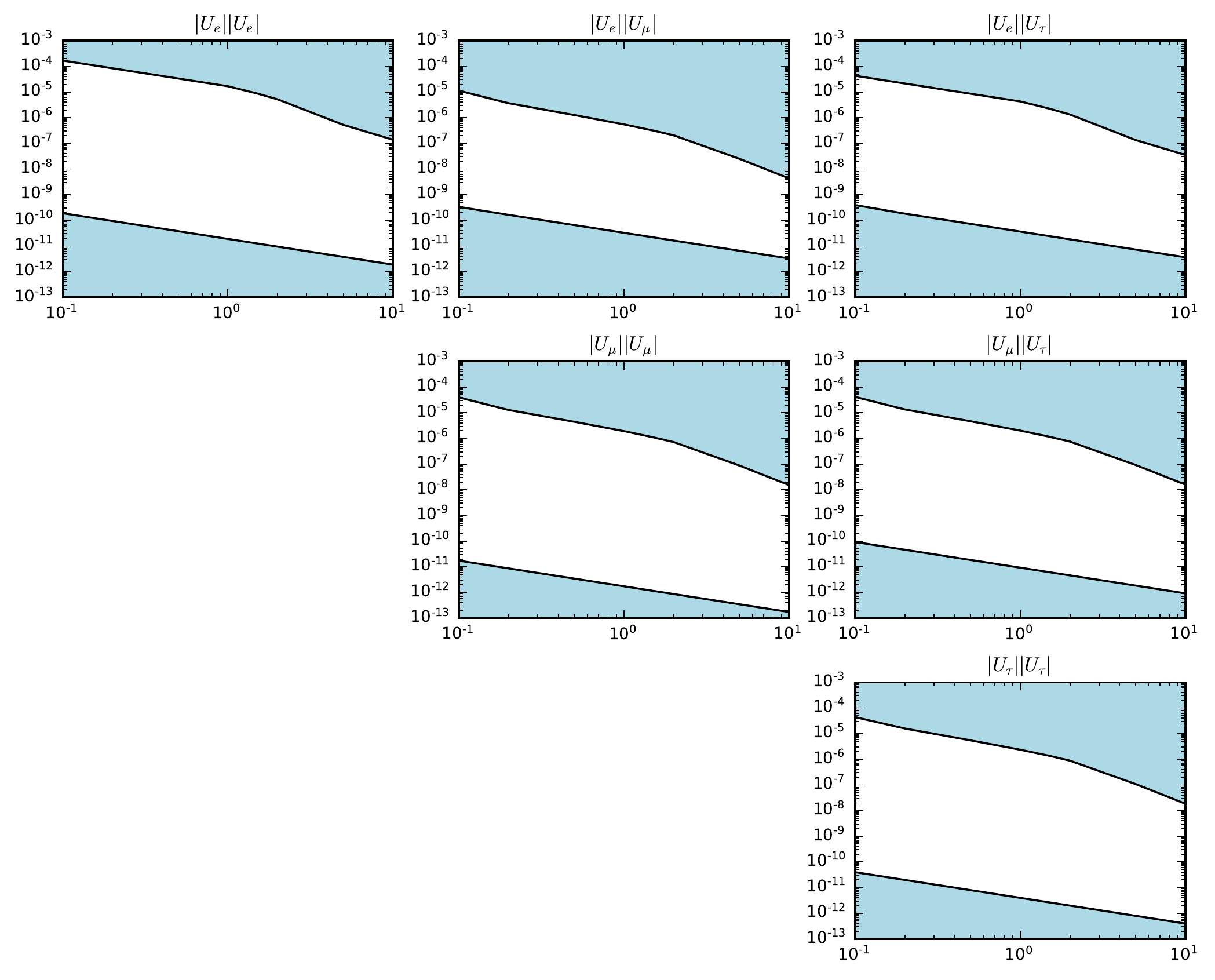}
\caption{\label{UaUb_IH} Cosmologically motivated regions of the individual mixings $|U_\alpha|^2$
and products $|U_\alpha|\cdot|U_\beta|$. Within the white regions it is possible to reproduced the observed
value of the BAU. The common mass is shown in the horizontal
axes, whereas the vertical axes show the corresponding product of mixings.
IH case.}
\end{figure}
}


\section{Open access datasets} 
\label{sec:open_access_data_sets}

In the previous section we have presented the boundaries of the regions where
all observed BAU can be addresses within the $\nu$MSM in terms of  
various combinations of the mixings of HNLs and active neutrinos.
However, the parameter space of the $\nu$MSM is not completely determined by the
plots presented above, and there are more hidden parameters. These parameters
can be essential for experimental searches for different signatures, and, e.g.
it is interesting to know branching ratios, such as $N\to \pi \ell_\alpha$
determined by $U_e:U_\mu:U_\tau$.
 For instance, what ratios of $U_e:U_\mu:U_\tau$
are possible for some point in the allowed region of
figure~\ref{UaUb_NH} or~\ref{UaUb_IH}? This information is crucial
to determine the decay length and branching ratios of various detection channels.
In order to fill in this gap, we publish several datasets~\cite{IT}.
\begin{itemize}
	\item Upper and lower boundaries of the  $M - |U|^2$ region where successful baryogenesis is possible as functions of the common mass of HNLs.
	Note that these lines correspond to the region where one can obtain $Y_B\ge Y_B^{obs}$.
	\item The  dataset of models with successful baryogenesis.
	The value of the BAU for every model (a parameter set) in this list 
	lies in the range $[Y_B^{obs}/2, 2\cdot Y_B^{obs}]$. Note that the value
	of the $Y_B$ is recorded for each parameter set so one can easily perform 
	another cuts.
	The models of this dataset can be used to perform 
	detailed Monte Carlo simulations of the experiments because
	they contain all necessary information ($M, |U_e|^2, |U_\mu|^2, |U_\tau|^2$).
	\item The dataset of models leading to various values of the BAU.
	Even though not all of these models provide a correct value of the BAU, they
	can be used to compare different theoretical approaches.
	\item Selected benchmark points are gathered in 
	appendix~\ref{sec:benchmark_points}.
\end{itemize}


\section{Generation of the baryon asymmetry} 
\label{sec:production_of_the_baryon_asymmetry}
\subsection{Kinetic equations} 
\label{sub:kinetic_equations}

In this section, we discuss the machinery of baryogenesis in the $\nu$MSM.
We present the kinetic equations which form the basis of the numerical analysis of this paper.
These equation possess the same generic structure as those used in 
refs.~\cite{Asaka:2005pn,Shaposhnikov:2008pf,Canetti:2012kh,Drewes:2016gmt,Hernandez:2016kel,Drewes:2016jae}.
However, several important improvements are incorporated. These are:
\emph{(i)} splitting of the rates to fermion number conserving and fermion number violating ones~\cite{Eijima:2017anv,Ghiglieri:2017gjz}; \emph{(ii)}
accounting for neutrality of the electroweak plasma 
(this requirement was added to kinetic equations in  \cite{Shuve:2014zua}) and  \emph{(iii)}
non-instantaneous freeze-out of sphalerons studied in~\cite{Eijima:2017cxr,Ghiglieri:2017csp}.
The rates entering the kinetic equations are updated using the recent results of ref.~\cite{Ghiglieri:2018wbs}.
The only remaining source of possible uncertainties is the averaging procedure described below.


The detailed derivation of the equations is presented in 
appendix~\ref{sec:derivation_of_kinetic_equations}. 
Here we start from the system of kinetic equations, introduce an ansatz which allows us
to integrate these equations over the momentum and show how the gradual 
freeze-out of the sphaleron processes can be accounted for.  The subscripts $2$ and $3$ are inherited from the $\nu$MSM and used to distinguish two HNLs participating
in the generation of the BAU.

We are interested in coherent oscillations of HNLs and their interactions with leptons.
The HNLs $N_2$ and $N_3$ are Majorana fermions with two helicity states each.
Helicities are used to distinguish particles from anti-particles. 
We assign positive fermion number to HNLs with positive helicity and vice versa.
Distribution functions and correlations of two HNLs are 
combined into matrices of density $\rho_N$ ($\rho_{\bar{N}}$ for antiparticles).
The kinetic equations for leptons are presented in terms  of the densities
of the $\Delta_\alpha=L_\alpha-B/3$, where $L_\alpha$ are the lepton numbers and
$B$ is the total baryon number. These combinations are not affected by the fast 
sphaleron processes and change only due to interactions with HNLs, therefore 
their derivatives are equal to the derivatives of the lepton number densities $n_{L_\alpha}$.
Here we present the equations determining the generation of asymmetries in terms of
$\Delta_\alpha$. In the next subsection, these asymmetries are related
to the BAU.
The system of kinetic equations reads 
\begin{subequations}
\begin{align}
i \frac{d n_{\Delta_\alpha}}{dt}
&= -  2 i \frac{\mu_\alpha}{T} \int \frac{d^{3}k}{(2 \pi)^{3}} \Gamma_{\nu_\alpha} f_{\nu} (1-f_{\nu})  \, 
    + i \int \frac{d^{3}k}{(2 \pi)^{3}} \left( \, \text{\text{Tr}}[\tilde{\Gamma}_{\nu_\alpha} \, \rho_{\bar{N}}]
    -  \, \text{\text{Tr}}[\tilde{\Gamma}_{\nu_\alpha}^\ast \, \rho_{N}] \right),\label{kin_eq_a}
\\
i \, \frac{d\rho_{N}}{dt} 
&= [H_N, \rho_N]
    - \frac{i}{2} \, \{ \Gamma_{N} , \rho_{N} - \rho_N^{eq} \} 
    - \frac{i}{2} \, \sum_\alpha \tilde{\Gamma}_{N}^\alpha \, \left[ 2 \frac{\mu_\alpha}{T} f_{\nu} (1-f_{\nu}) \right],\label{kin_eq_b}
\\
i \, \frac{d\rho_{\bar{N}}}{dt} 
&= [H_N^\ast, \rho_{\bar{N}}]
    - \frac{i}{2} \, \{ \Gamma_{N}^\ast , \rho_{\bar{N}} - \rho_N^{eq} \} 
    + \frac{i}{2} \, \sum_\alpha (\tilde{\Gamma}_{N}^\alpha)^\ast \, \left[ 2 \frac{\mu_\alpha}{T} f_{\nu} (1-f_{\nu}) \right].
\label{kin_eq_c}
\end{align}\label{KE_1a}\end{subequations}
In~\eqref{KE_1a} $f_\nu = 1/\left( e^{k/T}+1 \right) $ is the Fermi-Dirac
distribution function of a massless neutrino.
The effective Hamiltonian describing the 
coherent oscillations of HNLs is
\begin{equation}
 H_N = H_0 + H_I, \quad
 H_0 = -\frac{\Delta M M}{E_N} \sigma_{1}, \quad
 H_I = h_{+} \sum_{\alpha} Y_{+, \alpha}^{N} + h_{-} \sum_{\alpha} Y_{-, \alpha}^{N},
\label{H_N}
\end{equation}
where $E_{N}=\sqrt{k_{N}^{2} + M^{2}}$ and $\sigma_1$ is the first Pauli matrix.

The damping rates are
\begin{equation}
	\begin{aligned}
	\Gamma_N &= \Gamma_+ + \Gamma_-,\quad
	\Gamma_+ = \gamma_+ \sum_{\alpha} Y_{+, \alpha}^{N},\quad
	\Gamma_- = \gamma_- \sum_{\alpha} Y_{-, \alpha}^{N},\\
	\Gamma_{\nu_\alpha} &= (\gamma_+ + \gamma_-) \sum_I h_{\alpha I} h_{\alpha I}^\ast.
	\end{aligned}
\label{damping}
\end{equation}
The communication terms, describing the transitions from HNLs to active neutrinos, are
\begin{equation}
	\tilde{\Gamma}_N^\alpha = - \gamma_{+} Y_{+, \alpha}^{N} + \gamma_{-} Y_{-, \alpha}^{N}, \quad 
	\tilde{\Gamma}_{\nu_\alpha} = - \gamma_+  Y_{+,\alpha}^{\nu}+ \gamma_- Y_{-,\alpha}^{\nu}.
	\label{communication}
\end{equation}
In the expressions above the subscripts $+$ and $-$
refer to the fermion number conserving and violating quantities correspondingly. 
The functions $h_\pm$ and $\gamma_\pm$
depend only on kinematics (i.e. on the
common mass of HNLs). These functions have to be determined over the whole temperature region of the interest.
This region includes both symmetric and Higgs phases. In the Higgs phase, the rates can be split---in terms of ref.~\cite{Ghiglieri:2016xye}---into ``direct'' and ``indirect'' contributions\footnote{Note that this separation is gauge dependent~\cite{Ghiglieri:2016xye}. It is the sum of direct and indirect contributions which is gauge independent.}. The direct contributions correspond to the processes where the Higgs field actually propagates. These reactions are also present in the symmetric phase.
The processes with the Higgs field replaced by its temperature dependent expectation value give rise to the indirect contributions. These indirect contributions are crucial at low temperatures. As we show below, they are also important for the  baryogenesis.
\begin{equation}
\begin{aligned}
 h_+ &= h_+^{\rm direct} +  \frac{2 \langle \Phi \rangle^2 E_\nu (E_N + k) (E_N + E_\nu)}
            {k E_N \left( 4(E_N+E_\nu)^2 +\gamma_{\nu (+)}^2 \right) }, \\
 h_- &= h_-^{\rm direct} +\frac{2 \langle \Phi \rangle^2 E_\nu (E_N - k) (E_N - E_\nu)}
            {k E_N \left( 4(E_N-E_\nu)^2 +\gamma_{\nu (-)}^2 \right) }, \\
 \gamma_+ &= \gamma_+^{\rm direct} +\frac{2 \langle \Phi \rangle^2 E_\nu (E_N + k)\gamma_{\nu (+)} }
            {k E_N \left( 4(E_N+E_\nu)^2 +\gamma_{\nu (+)}^2 \right) }, \\
 \gamma_- &= \gamma_-^{\rm direct} + \frac{2 \langle \Phi \rangle^2 E_\nu (E_N - k)\gamma_{\nu (-)}  }
            {k E_N \left( 4(E_N-E_\nu)^2 +\gamma_{\nu (-)}^2 \right) },
\end{aligned}
\label{gamma_pm}
\end{equation}
In eqs.~\eqref{gamma_pm} the first and second terms represent the direct and indirect contributions respectively.
Below we discuss the direct contributions, the neutrino dumping rates $\gamma_{\nu, (\pm)}$, and
the neutrino potential in medium $b$ entering eq.~\eqref{gamma_pm} through $E_\nu = k - b$. The derivation of the indirect contributions is presented in Appendix~\ref{sec:derivation_of_kinetic_equations}.

The direct contributions to the effective Hamiltonians $h_\pm$ come from the real part of 
HNL's self energy $\slashed{\Sigma}_N(p) = \slashed{p} \alpha + \slashed{u} \beta$. Namely, one 
has\footnote{Note that the term proportional to $\alpha$ has been omitted in ref.~\cite{Ghiglieri:2018wbs} since it is subleading. Here we keep it for completeness. Note also that according to the formal 
power counting of ref.~\cite{Ghiglieri:2018wbs} the contribution $h_-^{\rm direct}$ has to be omitted as well.}
\begin{equation}
  h_\pm^{\rm direct} = \frac{1}{2p^0} \left( 
  \Re\beta (p^0 \pm p)  \mp  m_N^2\Re\alpha\right).
\end{equation}
At hight-temperature limit the function $h_+$ reproduces the standard Weldon correction $T^2/(8k)$~\cite{Weldon:1982bn}. If one neglects $\alpha$, which is numerically insignificant, the $h_-$ is suppressed compared to $h_+$ by a factor $M_N^2/p^2$. Thus $h_-$ is very small in the symmetric phase and, at the same time, the indirect contribution dominates in the Higgs phase. In our numerical computations we use the real part of the HNL safe energy calculated in ref.~\cite{Ghiglieri:2016xye}.

We now move to the direct contributions $\gamma_\pm^{\rm direct}$. They originate from $1\leftrightarrow2$,
$2\leftrightarrow2$, and $1 + n \leftrightarrow 2 + n$ processes. The latter two require proper 
resummations~\cite{Anisimov:2010gy,Ghiglieri:2017gjz}.
Fermion number conserving rate comes mainly from the $2 \leftrightarrow 2 $ scatterings . Another contribution to $\gamma_+^{\rm direct}$ comes from $1 + n \leftrightarrow 2 + n$. 
Fermion number violating rate comes from the $1 + n \leftrightarrow 2 + n$ processes
(note that $2 \leftrightarrow 2 $ scatterings do not contribute to $\gamma_-^{\rm direct}$).
For our numerical analysis we use $\gamma_\pm^{\rm direct}$ kindly provided by the authors of ref.~\cite{Ghiglieri:2018wbs}. It is important to stress that in ref.~\cite{Ghiglieri:2018wbs} the temperature of the electroweak crossover $T_c$ has been extracted from the one loop correction to the Higgs potential. Computed this way, $T_c \simeq 150$~GeV~\cite{Ghiglieri:2016xye}, whereas the non-perturbative result is 
$T_c^{\rm NP} \simeq 160$~GeV \cite{Laine:2015kra,DOnofrio:2015gop}. Since we are using the rates from ref.~\cite{Ghiglieri:2018wbs}, the crossover temperature is set to  $T_c \simeq 150$~GeV. We also implement one-loop running of the couplings following the approach of ref.~\cite{Ghisoiu:2014ena}.

The last two ingredients entering \eqref{gamma_pm} are the neutrino dumping rates $\gamma_{\nu, (\pm)}$  and the neutrino potential in the medium $b$.
The function $b$ can be calculated following, e.g. refs.~\cite{Notzold:1987ik,Morales:1999ia}.
The neutrino damping rates are related to its self energy
$\slashed{\Sigma}_\nu(k) = \slashed{k} \left( a+ i \Gamma_k/2 \right) 
  + \slashed{u} \left( b + i \Gamma_u/2 \right)$ as
\begin{equation}
  \gamma_{\nu (+)} = \Gamma_u + 2 k^0 \Gamma_k, \quad \gamma_{\nu (-)} = \Gamma_u.
  \label{two_nu_rates}
\end{equation}
In the temperature region of interest, the neutrino damping rates are dominated
by $2 \leftrightarrow 2$ process mediated by soft gauge bosons. The calculation 
of these rates requires proper resummations~\cite{Ghiglieri:2016xye}. We use analytical results presented in 
ref.~\cite{Ghiglieri:2018wbs}.

The dependence on the Yukawa coupling constants factorises out
from the rates~\eqref{H_N}--\eqref{communication}.
It is convenient to introduce the matrix of Yukawa couplings $h_{\alpha I}$ related to the matrix
$F_{\alpha I}$ defined in~\eqref{Lagr} as follows
\begin{equation}
F_{\alpha I} = h_{\alpha J} [U_N^\ast]_{JI}, \quad
U_N = \frac{1}{\sqrt{2}}\begin{pmatrix} -i & 1 \\ i & 1 \end{pmatrix}.
\end{equation}
In terms of these couplings we have
\begin{equation}
\begin{aligned}
Y_{+, \alpha}^{N} &= \left( Y_{+, \alpha}^{\nu} \right)^T = 
\begin{pmatrix}
  h_{\alpha 3} h_{\alpha 3}^\ast & - h_{\alpha 3} h_{\alpha 2}^\ast \\
 - h_{\alpha 2} h_{\alpha 3}^\ast & h_{\alpha 2} h_{\alpha 2}^\ast
\end{pmatrix},\\
Y_{-, \alpha}^{N} &= \left( Y_{-, \alpha}^{\nu} \right)^T = 
\begin{pmatrix}
  h_{\alpha 2} h_{\alpha 2}^\ast & - h_{\alpha 3} h_{\alpha 2}^\ast \\
 - h_{\alpha 2} h_{\alpha 3}^\ast & h_{\alpha 3} h_{\alpha 3}^\ast
\end{pmatrix}.
\end{aligned}
\end{equation}

The relation between the number densities and the chemical potentials to leptons
in eq.~\eqref{KE_1a}
has to take into account the neutrality of plasma.
When the system is in equilibrium with respect to sphaleron processes,
this relation reads
\begin{equation}
	\mu_\alpha = \omega_{\alpha \beta}(T) n_{\Delta_\beta},
	\label{susceptibilities}
\end{equation}
where $\omega_{\alpha \beta}(T)$ is the so-called susceptibility matrix,
see, e.g.~\cite{Ghiglieri:2016xye,Eijima:2017cxr}. In the symmetric phase
its diagonal elements are ${\omega_{\alpha \alpha} = 514/(237 \, T^2)}$,
 while
the off-diagonal $\omega_{\alpha \beta} = 40/(237 \, T^2), \alpha \neq \beta$.
Note that relation~\eqref{susceptibilities} should be modified for
temperatures below the $T_{sph}\simeq  131.7$~GeV
at which the sphalerons decouple~\cite{DOnofrio:2014rug}.
Full expressions of the susceptibility matrices can be found in ref.~\cite{Eijima:2017cxr}.

The set~\eqref{KE_1a} is 
a system of coupled integro-differential equations for each momentum mode of HNLs.
Numerical solution of this system is a very complicated 
task~\cite{Asaka:2011wq,Ghiglieri:2017csp}.
However, a certain ansatz could be made to simplify the system.
Namely, let us assume that the momentum dependence of the distribution
functions is the equilibrium one, $\rho_{X} (k,t) = R_{X}(t) f_N(k)$, where
$f_N(k)$ is the Fermi-Dirac distribution of the massive HNLs.
Then it is possible to integrate the kinetic
equations over the momentum and obtain a set of ordinary differential equations.
This procedure is the main source of the theoretical uncertainty.
The error can be estimated by comparing solutions of the averaged
equations with solutions of the full set~\eqref{KE_1a}.
This has been done first in  ref.~\cite{Asaka:2011wq}. Results of this work indicate that the error in the value of the BAU doesn't
exceed $40 \%$. Authors of the recent study~\cite{Ghiglieri:2017csp} 
have also solved the full system. They have found that the accurate result
differs by a factor of $1.5$ from the benchmark of ref.~\cite{Hernandez:2016kel}.
However, note that the equations of ref.~\cite{Ghiglieri:2017csp} include effects
that haven't been accounted for in ref.~\cite{Hernandez:2016kel}.
We have also tested the benchmark points listed in~\cite{Mikko}
using the same neutrino oscillation data as in~\cite{Ghiglieri:2017csp} and
found a surprisingly good agreement for the most of the benchmark points.
A file with the values of the BAU for all benchmark points could be found 
in~\cite{IT}.
In what follows we will be conservative and assume that the 
averaging can lead to a factor of two error.

Let us finally present the system of equations that we actually solve numerically.
It is  convenient to introduce the CP-even and CP-odd combinations
$\rho_+ \equiv (\rho_N+\rho_{\bar{N}})/2 - \rho_N^{eq}$,\quad  
$\rho_- \equiv \rho_N - \rho_{\bar{N}}$.
The averaged equations read
\begin{equation}
\begin{aligned}
 \dot{n}_{\Delta_{\alpha}} = &- \text{Re}\,\overline{\Gamma}_{\nu_{\alpha}} 
 \mu_\alpha\frac{T^2}{6} 
 + 2 i \Tr [(\text{Im}\,\overline{\tilde{\Gamma}}_{\nu_{\alpha}})n_{+}] 
 - \Tr [(\text{Re}\,\overline{\tilde{\Gamma}}_{\nu_{\alpha}}) n_{-}], \\
 \dot{n}_{+} = &-i [\text{Re}\,\overline{H}_N,n_{+}] + \frac{1}{2}[\text{Im}\,\overline{H}_N,n_{-}] 
 - \frac{1}{2}\{\text{Re}\,\overline{\Gamma}_N,n_{+}\} - \frac{i}{4} \{\text{Im}\,\overline{\Gamma}_N,n_{-}\} \\
 &  - \frac{i}{2} \sum (\text{Im}\,\overline{\tilde{\Gamma}}_N^{\alpha})\mu_\alpha\frac{T^2}{6}- S^{eq}, \\ 
 \dot{n}_{-} =&\; 2 [\text{Im}\,\overline{H}_N,n_{+}] -i [\text{Re}\,\overline{H}_N,n_{-}]
  - i\{\text{Im}\,\overline{\Gamma}_N,n_{+}\} - \frac{1}{2} \{\text{Re}\,\overline{\Gamma}_N,n_{-}\} \\
  &- \sum (\text{Re}\,\overline{\tilde{\Gamma}}_N^{\alpha}) \mu_\alpha\frac{T^2}{6},
 \label{KE_2}
\end{aligned}
\end{equation}
with the integrated rates defined as
\begin{equation}
\begin{aligned}
 \overline{\Gamma}_{\nu_{\alpha}} &= \frac{6}{\pi^2} \int dk_c k_c^2 e^{k_c} f_\nu^2 \Gamma_{\nu_\alpha}, \quad
 \overline{\tilde{\Gamma}}_{\nu_{\alpha}} = \frac{T^3}{2 \pi^2} \frac{1}{n_N^{eq}} \int dk_c k_c^2 f_N \tilde{\Gamma}_{\nu_\alpha}, \\
 \overline{H}_N &= \frac{T^3}{2 \pi^2} \frac{1}{n_N^{eq}} \int dk_c k_c^2 f_N H_N, \quad
 \overline{\Gamma}_N = \frac{T^3}{2 \pi^2} \frac{1}{n_N^{eq}} \int dk_c k_c^2 f_N \Gamma_N, \\
 \overline{\tilde{\Gamma}}^{\alpha}_N &= \frac{6}{\pi^2} \int dk_c k_c^2 e^{k_c} f_\nu^2 \tilde{\Gamma}^{\alpha}_N, \quad
 S^{eq} = \frac{T^3}{2 \pi^2} \frac{1}{s} \int dk_c k_c^2 \dot{f_N}\cdot
 {\bf 1}_{2\times 2}.
\end{aligned}
\label{integrated_rates}
\end{equation}

Equations~\eqref{KE_2}  are formulated in a static universe.
The expansion of the Universe can be accounted for by rewriting
eqs~\eqref{KE_2} in terms of the so-called yields $Y_X = n_X/s$, where 
$n_X$ is the number density of species $X$ and $s$
is the entropy  density which is conserved in the co-moving volume.
For the numerical computations
we use $s(T)$ calculated in refs.~\cite{Laine:2006cp,Laine:2015kra}.

\subsection{Gradual freeze-out of sphalerons} 
\label{sub:gradual_freeze_out_of_sphalerons}

The asymmetry generated in the lepton sector is communicated to the baryon
sector by the sphaleron processes. As long as these processes are fast compared to the
rate of the asymmetry generation, the following equilibrium relation 
holds~\cite{Khlebnikov:1996vj,Burnier:2005hp}
\begin{equation}
	Y_{B^{eq}}=- \chi(T)  \sum_\alpha Y_{\Delta_\alpha} , \quad \chi(T) =
	 \frac{4 \left(27 (\sqrt{2}\langle \Phi \rangle/T)^2+77\right)}{333 
	 (\sqrt{2}\langle \Phi \rangle/T)^2+869},
	 \label{B_eq}
\end{equation}
where  $\langle \Phi \rangle$ is the Higgs vacuum expectation value, 
which is equal to $174.1$~GeV at zero temperature.
However, as was demonstrated
in ref.~\cite{Eijima:2017cxr}, a deviation from the equilibrium with respect
to sphalerons happens at temperatures around $140$~GeV, i.e., before the freeze-out.
It was also shown that the errors stemming from the usage of the 
equilibrium formula can exceed an order of magnitude. 
To overcome this problem, one can implement the method suggested in 
ref.~\cite{Eijima:2017cxr}.
Namely, one solves the kinetic equation for the baryon 
number~\cite{Khlebnikov:1988sr,Burnier:2005hp}
\begin{equation}
	\dot{Y}_B = -\Gamma_B (Y_B-Y_{B^{eq}}),
	\label{kine_ec_B}
\end{equation}
where for the three SM generations
\begin{equation}
   \Gamma_B= 3^2\cdot \frac{869 + 333 (\sqrt{2}\langle \Phi \rangle/T)^2}{792 +
   306 (\sqrt{2}\langle \Phi \rangle/T)^2}
    \cdot \frac{\Gamma_{diff}(T)}{T^3} , 
   \label{Gamma_B}
\end{equation}
were $Y_{B^{eq}}$ is given by eq.~\eqref{B_eq}
and  $\sum_\alpha Y_{\Delta_\alpha}$ is calculated from the main system~\eqref{KE_2}.
It is enough to solve eq~\eqref{kine_ec_B} starting from $T=150$~GeV.

This finishes the presentation of the kinetic equations. To summarize, our 
equations incorporate all physical effects that are relevant
for the  range of HNL masses  considered here. The only source of errors is the momentum averaging of 
the kinetic equations. Based on the previous studies~\cite{Asaka:2011wq,Ghiglieri:2017csp}, we conservatively estimate that these errors do not
exceed a factor of two. 


\subsection{Physics of asymmetry generation } 
\label{sub:physics_of_asymmetry_generation_}

Before solving eqs.\eqref{KE_2} numerically, we briefly discuss the physics of 
the asymmetry generation.
There are several important temperature scales. 
One of them is the already mentioned sphaleron freeze-out temperature
$T_{sph}\simeq  131.7$~GeV~\cite{DOnofrio:2014rug}.
The lepton asymmetry generated at temperatures lower than $T_{sph}$ does
not affect the final value of $Y_B$.

Second  scale is a temperature at which the HNLs enter thermal equilibrium, 
 $T_{in}$. This temperature can be determined from the condition
$\overline{\Gamma}_N(T_{in})/H(T_{in})\simeq 1$, where 
$H(T)$ is the Hubble rate and we take the largest eigenvalue
of the  matrix $\overline{\Gamma}_N(T_{in})$. The Hubble rate during
radiation-dominated epoch is $H(T) = T^2/M_{Pl}^*$, where 
$M_{Pl}^* = \sqrt{90/(8\pi^3 g_*)} \,M_{Pl}$, $g_*$~is the effective number 
of relativistic degrees of freedom.

Another important scale is the so-called oscillation temperature, $T_{osc}$.
Coherent oscillations of the HNLs play the crucial role in the generation of the individual
lepton asymmetries. In fact, they provide a CP-even phase, which, being combined with
the CP-odd phase from Yukawas, leads to the generation of the individual 
asymmetries, see, e.g. the discussion in ref.~\cite{Shaposhnikov:2008pf}.
This mechanism becomes efficient around the first oscillation.
The temperature at which the first oscillation takes place can be estimated as
\cite{Akhmedov:1998qx,Asaka:2005pn}
\begin{equation}
	T_{osc} \simeq \left( \frac{\delta M M M_{Pl}^*}{3} \right)^{1/3},
	\label{Tosc}
\end{equation}
where  $\delta M$ is the physical mass difference.
This mass difference is the splitting between two eigenvalues of the 
effective Hamiltonian~\eqref{H_N}.
Since the asymmetry generation is efficient at $T $ around $ T_{osc}$, one can roughly
estimate the lower bound on the $\delta M$ by requiring $T_{osc} > T_{sph}$.

Let us consider two cases:
\begin{itemize}
	\item $T_{in} < T_{Sph}$

	This regime is sometimes referred to as \emph{oscillatory}. 
	In this case  the kinetic equations could be solved 
	perturbatively~\cite{Akhmedov:1998qx,Asaka:2005pn}
	if also $T_{osc}>T_{Sph}$.\footnote{Note that when the fermion number violating effects are introduced, the total asymmetry is generated at order $F^4$.}
	Late thermalization implies that Yukawa couplings are relatively small. 
	In terms of the parameters listed in table~\ref{table_parameters},
	this regime is realized if the value of $|\imw|$ is small.

	\item $T_{in} \ge T_{Sph}$

	In this case, the dampings of the generated asymmetries are efficient, so this 
	scenario is referred to as \emph{strong wash-out regime}.
	Yukawa couplings must be relatively large, this can be in agreement with the oscillation data if $|\imw|$ is large as well.
	Sizeable damping causes a wash-out of asymmetries before the freeze-out of sphaleron processes. However, the production of HNLs and interactions with left-handed leptons at high temperatures are also enhanced, so the asymmetry generation is more efficient.
    In order to account for  all relevant processes, the kinetic equations have to be solved numerically.
\end{itemize}


\section{Numerical analysis of the kinetic equations} 
\label{sec:numerical_solution_of_the_equations}

Equations~\eqref{KE_2} together with~\eqref{B_eq} or~\eqref{kine_ec_B}
allow one to determine the value of the BAU for each parameter set in 
table~\ref{table_parameters}. 
For most values of the parameters, the equations~\eqref{KE_2} 
have to be solved numerically.
In this section, we discuss the procedure of solving these equations
and demonstrate how the improvements in the equations influence the results.

\subsection{The procedure for numerical solution of the equations} 
\label{sub:the_procedure_for_numerical_solution_of_the_equations}

First of all, it is necessary to determine initial conditions. 
Right after the inflation the baryon and lepton numbers
of the Universe as well as number densities of HNLs are equal to 
zero.\footnote{The effect of the initial asymmetry in the HNL sector has been studied
in ref.~\cite{Asaka:2017rdj}.}
Thus we start from the vanishing $Y_- (T_{0})\equiv n_-(T_{0})/s(T_{0})$ 
and $Y_{\Delta_\alpha}(T_{0})$.
According to the definition of $\rho_+$, at initial stage 
$Y_+(T_{0}) = - n^{eq}(T_{0})/s(T_{0})$, where
$n^{eq}(T)$ is an equilibrium number density of a fermion with mass $M$.

The appropriate initial temperature can be specified on physical grounds.
As we have discussed, the asymmetry generation starts around the 
 temperature of the first oscillation, $T_{osc}$.
We have checked numerically that no significant asymmetry is generated at the temperature
$10\cdot T_{osc}$, i.e. much before the onset of oscillations.
Therefore we take $T_{0}=10\cdot T_{osc}$ if $10\cdot T_{osc} > 10^3$~GeV, 
or $T_0 = 10^3$~GeV otherwise.

Now, having set up the initial conditions, we can solve
equations~\eqref{KE_2}. It is convenient to implement them 
using  $z = \log(M/T)$ as a variable.
Even though the problem  is reduced to the solution of the set of 11 ordinary
differential equations (ODE) by means of averaging~\eqref{integrated_rates},
it still remains challenging.
The reason is that significantly different time scales are present in the 
system~\eqref{KE_2}. Appropriate stiff ODE solvers, such as LSODA~\cite{LSODE},
can handle our equations quite efficiently. However, the integration
time can be reduced further. Notice 
that the effective Hamiltonian entering  equations~\eqref{KE_1a} 
can be decomposed as $H_N = H_0 + H_I$, where 
$H_0 = -\Delta M M \sigma_{1}/E_N $. Therefore, we can move to the
`interaction picture' with respect to $H_0$.
After this transformation, the equations can be solved using
a non-stiff method. 

The final value of the BAU is founded by solving eq.~\eqref{kine_ec_B}.
This  ensures that the value of the BAU is not affected by the assumption
of an instantaneous freeze-out of sphalerons.

In order to find an appropriate method we have implemented equations~\eqref{KE_2}
in the Python programming language. The 
SciPy library~\cite{scipy} allows one to use several different ODE solvers.
We have found that the most efficient (in terms of the number of calls of the r.h.s.)
one for our purposes is the LSODE~\cite{LSODE}. 
The equations were then coded in the Fortran 77/95 along with the native Fortran 
implementation of the LSODE~\cite{netlib}. Note that
for the successful integration it is important
to carefully tune the parameters of the solver (such as absolute and relative tolerances for each variable).

We have also implemented the same system of kinetic equations in  Mathematica 
\cite{Mathematica}. This allowed us to validate the results obtained
using the Fortran code. However,
since the computation of r.h.s. takes much longer, the overall computation time 
is also very large.
The Fortran realization outperforms the Mathematica one by more than four orders
of magnitude.

The whole computation of the BAU for a single set of the parameters takes from
$\simeq 0.05$~sec in the oscillatory regime (approximately $|\imw| < 2$)
to $\simeq 1.0$~sec in the strong wash-out regime 
(approximately $|\rew| > 5.5$).
The very efficient numerical procedure allows performing
a comprehensive study of the parameter space.


\subsection{Analysis of the equations} 
\label{sub:analysis_of_the_equations}

\begin{figure}[htb!]
\centerline{
    \includegraphics[width = 0.5\textwidth]{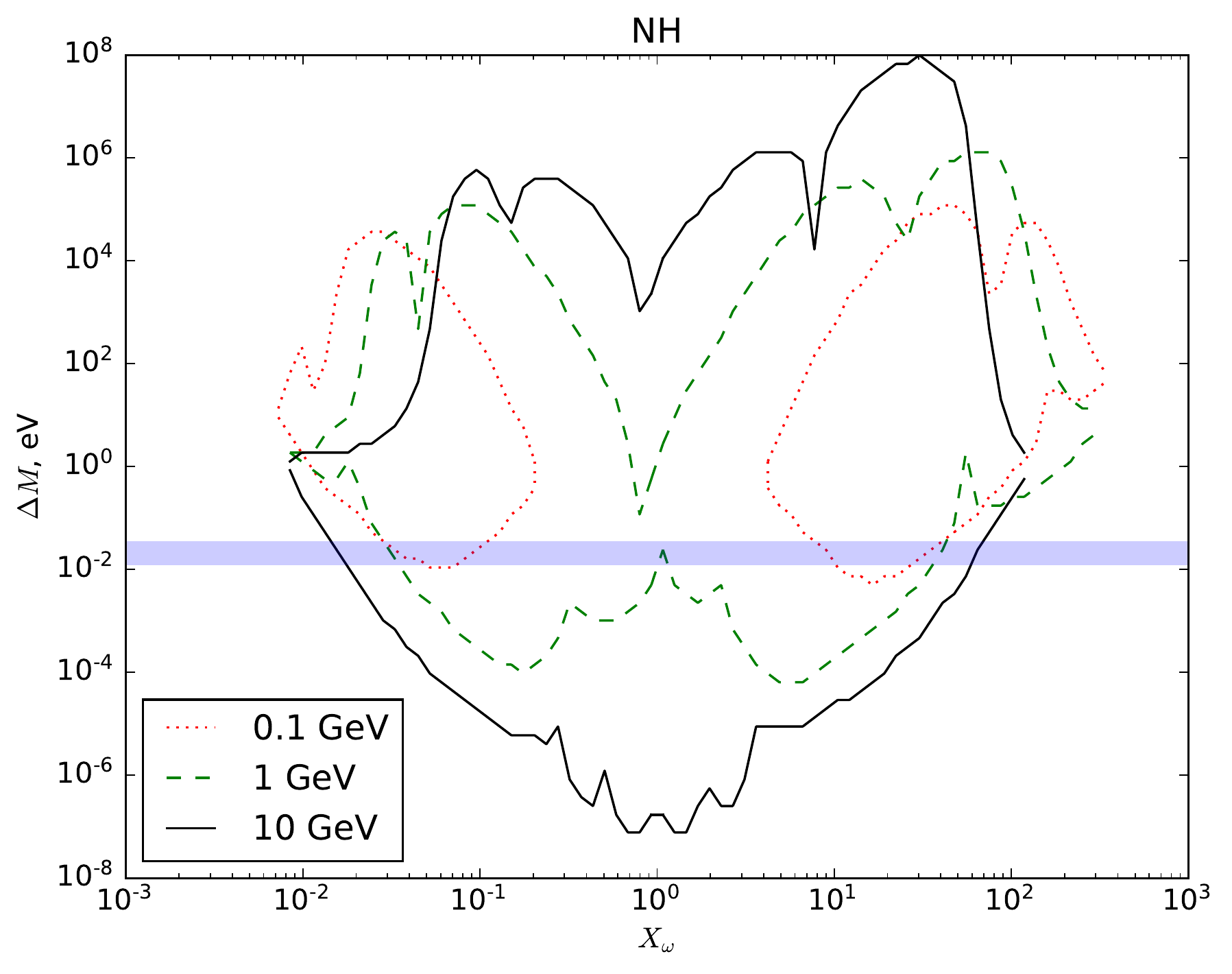}
    \includegraphics[width = 0.5\textwidth]{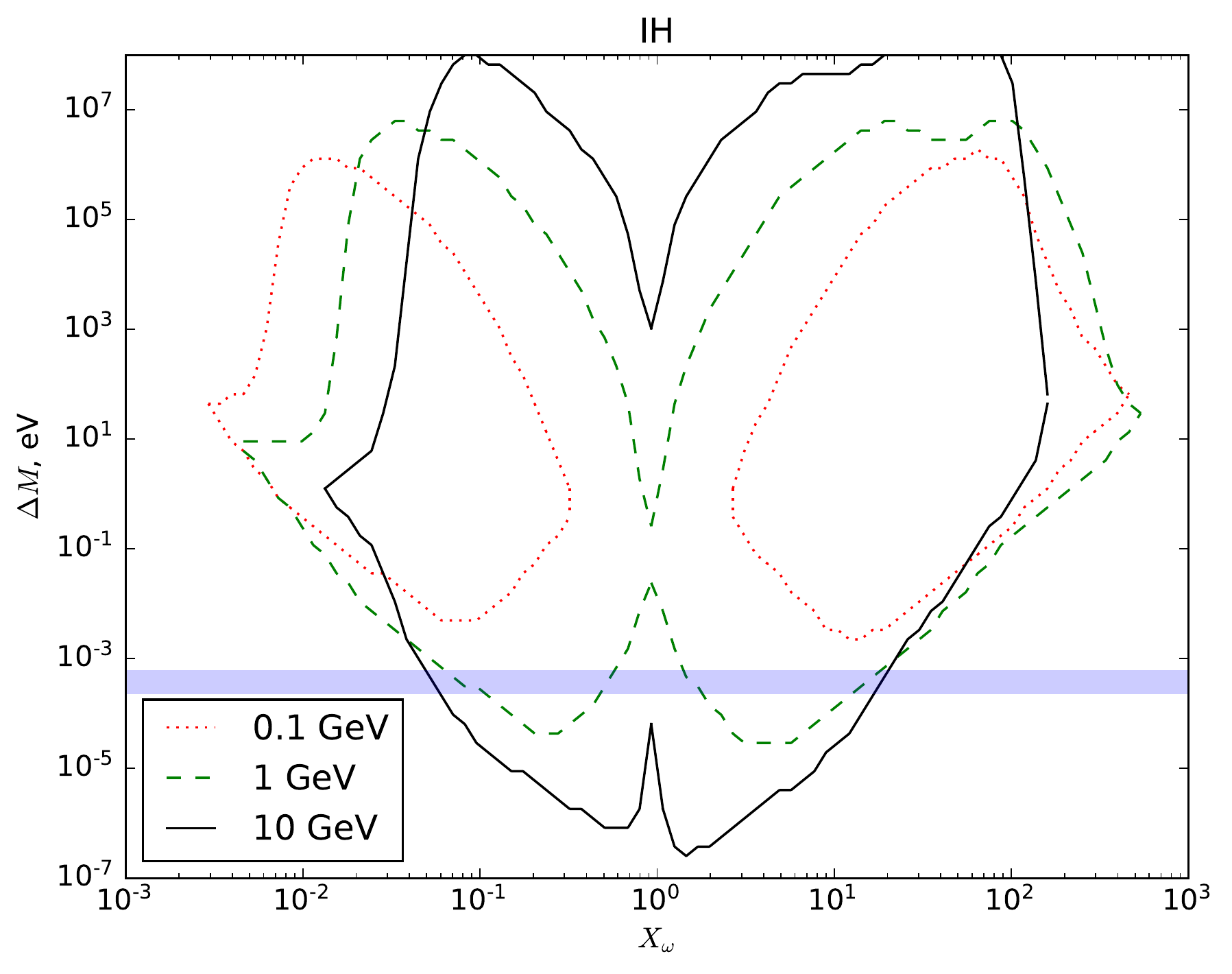}
}
\caption{\label{Xw_dM_bounds}Allowed region in the $X_\omega - \Delta M$ plane
obtained for the fixed values of the phases~\eqref{fixed_phases}.
It is possible to generate $Y_B\ge Y_B^{obs}$ within the corresponding regions.
Common masses of the HNLs are fixed to be equal to $0.1, 1.0, 10$~GeV.
The blue horizontal line indicates the zero-temperature Higgs contribution to the 
physical mass difference $\delta M$ in the limit $\Delta M/M\to0$. 
}
\end{figure}

Before  presenting the phenomenologically relevant results it is instructive
to study the outcome of the improvements of the kinetic equations.
First of all, it is interesting to see what values of mass splitting $\Delta M$
and $X_\omega$ can lead to a successful baryogenesis. 
 In figure~\ref{Xw_dM_bounds} we show the regions in the $X_\omega - \Delta M$ plane
where the value of the $Y_B\ge Y_B^{obs}$ can be generated.
To obtain this plot we have fixed the phases to the following values
\begin{subequations}
\begin{align}
	\mbox{NH:} &\quad \delta = \pi , \quad \eta = 3 \pi/2,\quad \rew = \pi/4,
    \label{NHphases0}\\
	\mbox{IH:} &\quad \delta =  0,\quad \, \eta = \pi/2,\quad \;\, \rew = \pi/4.
    \label{IHphases0}
\end{align}
\label{fixed_phases0}\end{subequations}
This choice of phases maximizes the value of $|Y_B|$ in the strong wash-out regime
(see more detailed discussion in section~\ref{sec:scan_of_the_parameter_spaces}).
Note that since the  values of Dirac and Majorana phases were fixed and only 
$\imw$ was varied, the contours are not symmetric.
In all cases the positive $\imw$ (large $X_\omega$) gives larger BAU,
as can be seen from figure~\ref{Xw_dM_bounds}.

The blue horizontal line in figure~\ref{Xw_dM_bounds}
indicates the zero-temperature Higgs contribution to the 
physical mass difference $\delta M$ in the limit $\Delta M/M\to0$. Below this line the Higgs contributions to the physical mass difference dominates,
whereas above the line, the  physical mass difference is mostly determined by the Majorana mass
difference. This means that $\delta M$ cannot be much smaller than $\Delta M$
in the region above the  line and $\delta M$ cannot be much smaller than
the Higgs contribution below the line.
Smaller values of $\delta M$---which are interesting, e.g. for studies
of resolvable HNL oscillations at the SHiP experiment---are only possible  if there is a cancellation
between $\Delta M$ and the Higgs contribution. This cancellation can happen only close to the blue line.

\begin{figure}[htb!]
\centerline{
    \includegraphics[width = 0.5\textwidth]{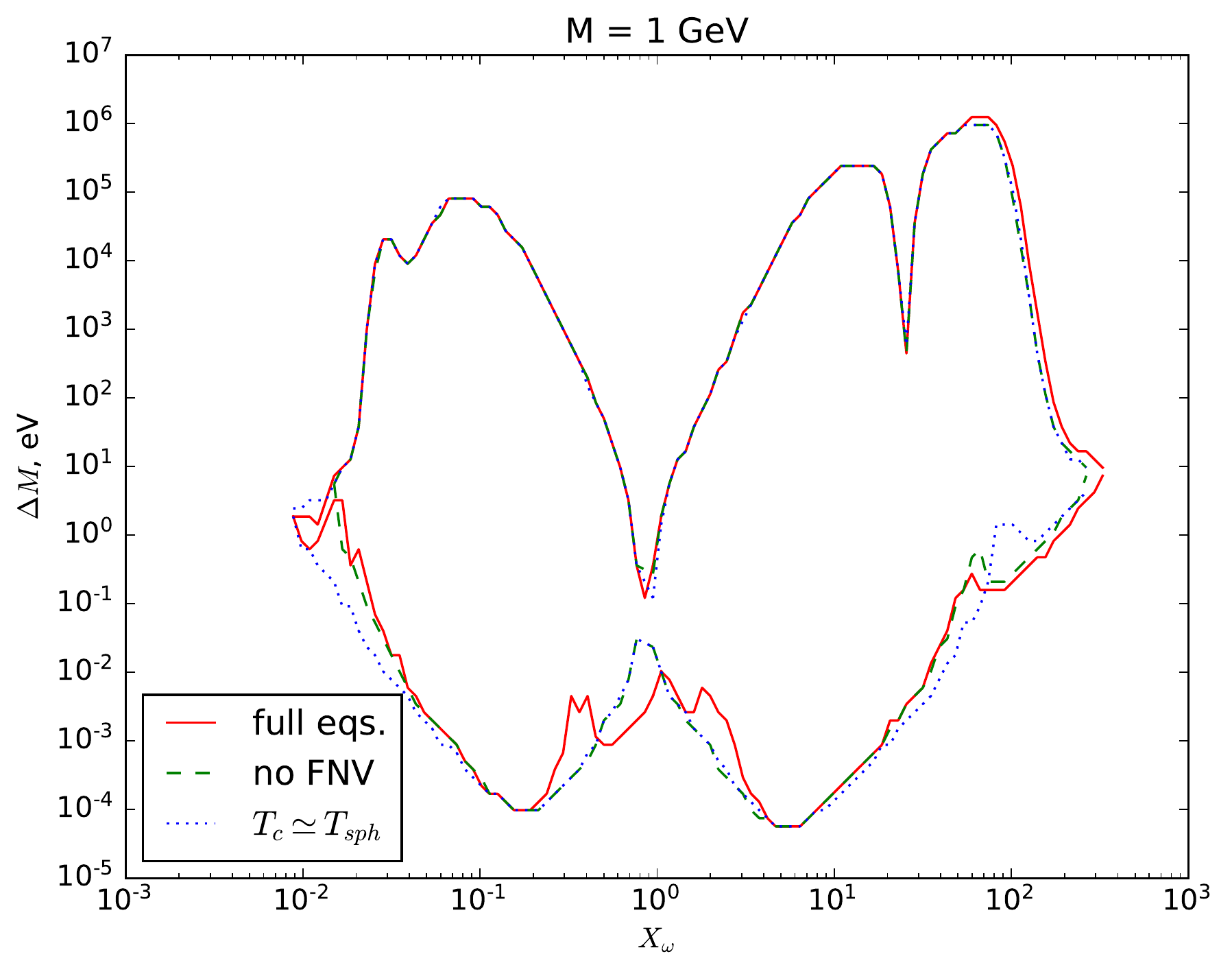}
    \includegraphics[width = 0.5\textwidth]{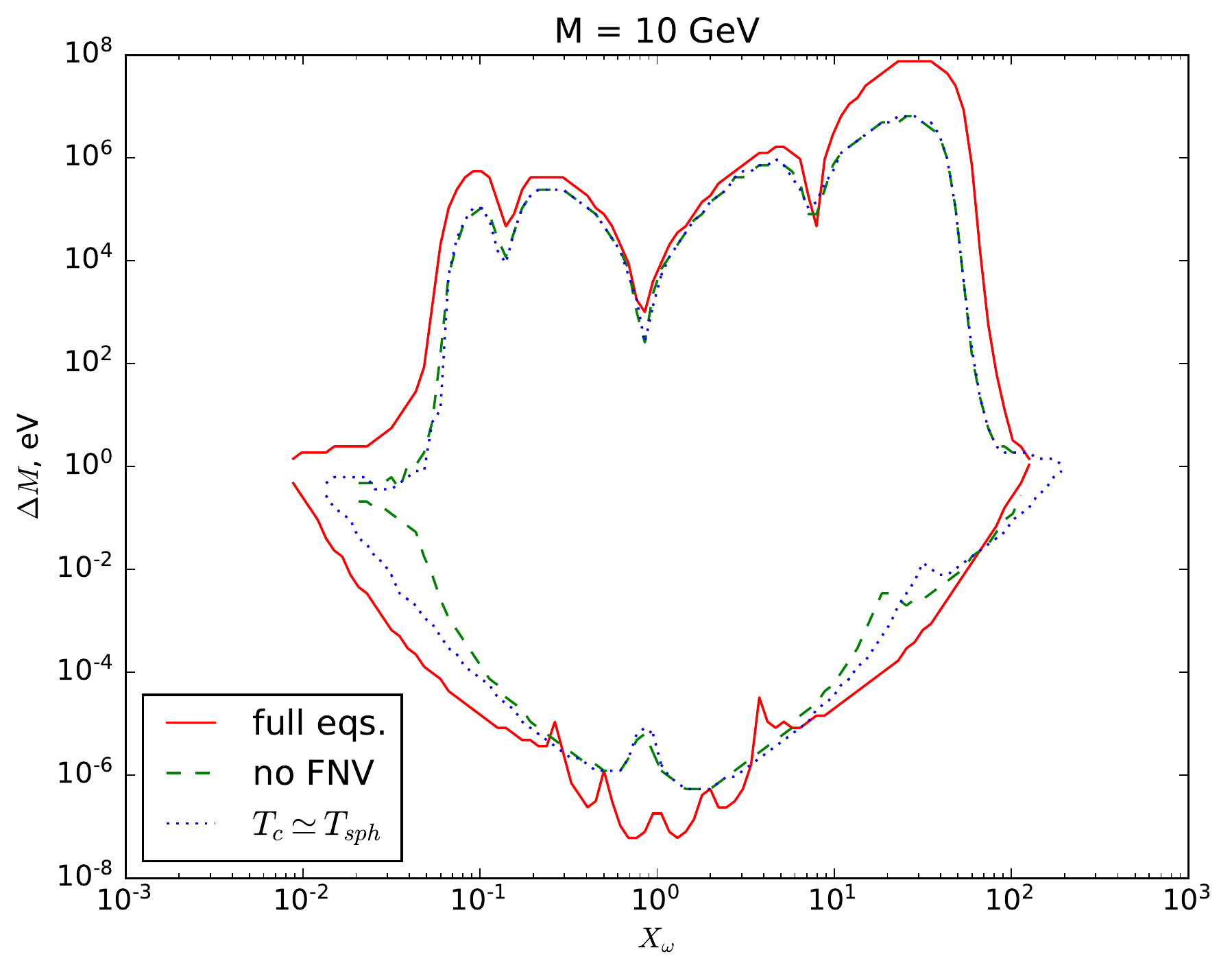}
}
\caption{\label{diff}
Red, solid line -- full kinetic equations~\eqref{KE_2}. 
No fermion number violation -- green, dashed lines.
Blue, dotted lines correspond to an assumption, $T_\text{EWPT} \simeq T_\text{sph}$, which has been used in a number of previous works so far.
\emph{Left panel}, common mass $M = 1$~GeV.
\emph{Right panel}, common mass $M = 10$~GeV.
}
\end{figure}

\begin{figure}[htb!]
\centerline{
    \includegraphics[width = 0.5\textwidth]{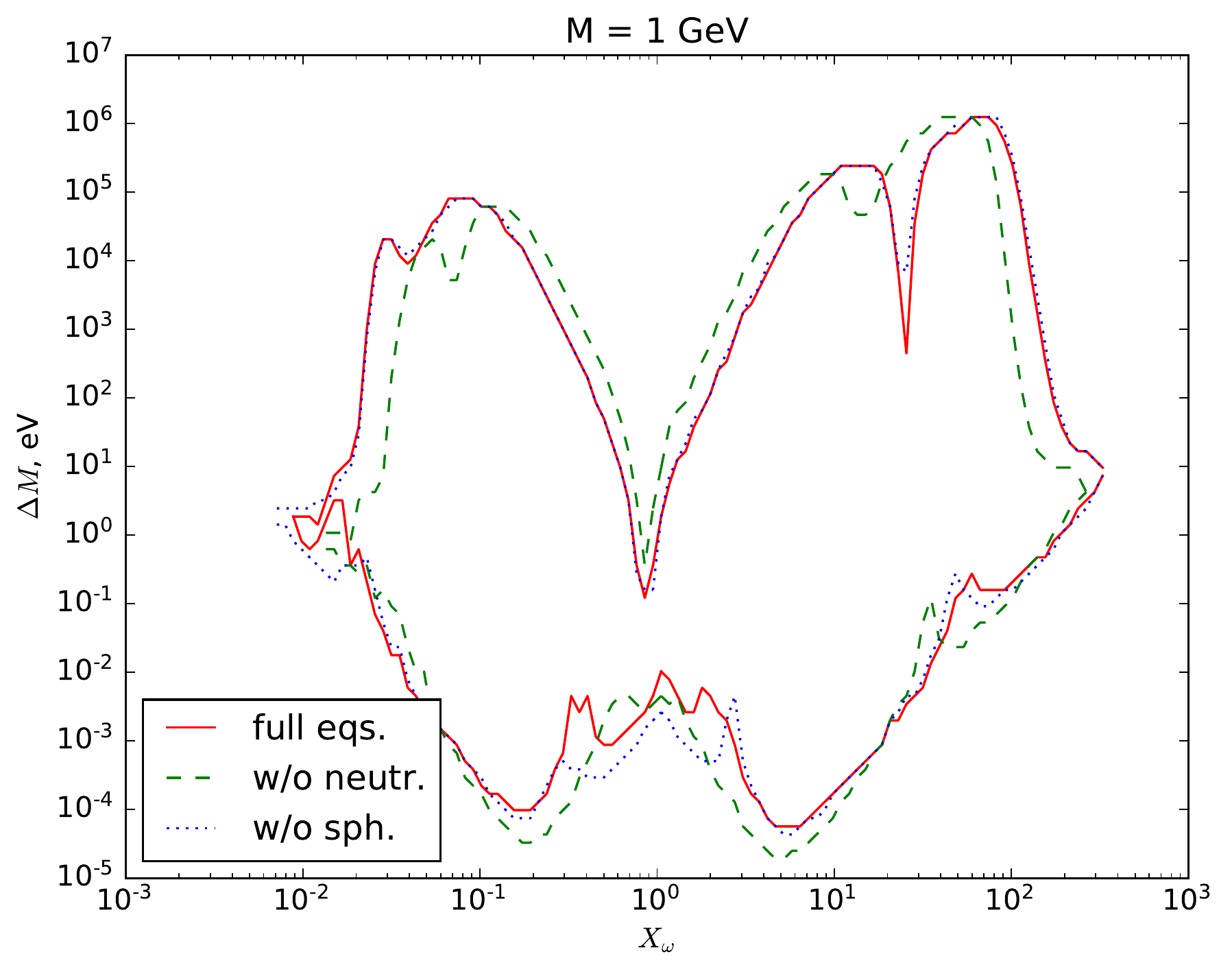}
    \includegraphics[width = 0.5\textwidth]{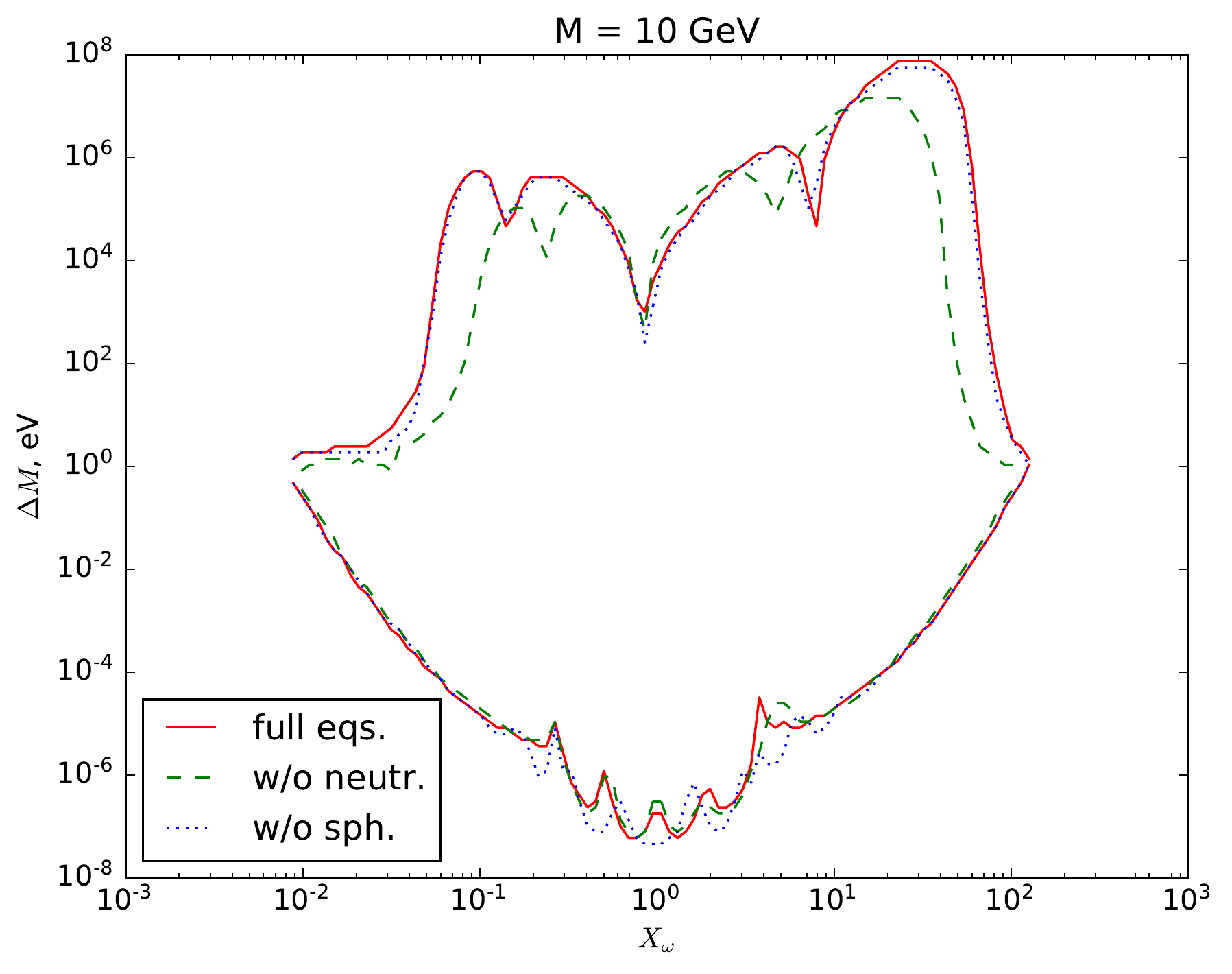}
}
\caption{\label{sus_sph}
Comparing the kinetic equations with accurate susceptibilities and accurate treatment 
of sphalerons (red curve), with diagonal susceptibilities (no plasma neutrality)
(green dashed curve) and with the instantaneous freeze-out of sphalerons (blue dotted curve). The same choice of phases as in the previous figure.
\emph{Left panel}, common mass $M = 1$~GeV.
\emph{Right panel}, common mass $M = 10$~GeV.
}
\end{figure}

It is also important to understand the role 
of the improvements that we consider. We want to address the questions: \emph{(i)} what is the effect of the fermion number
violating rates on the final value of the BAU; \emph{(ii)} what is the effect of
considering the Higgs phase;
\emph{(iii)} what is the effect of susceptibilities;
\emph{(iv)} what is the effect of the gradual decoupling of the sphalerons.
In order to answer the first question, we effectively switch off
fermion number violating processes in our kinetic equations. It is possible because
the fermion number conserving and fermion number violating processes are neatly
separated in eqs.\eqref{H_N},~\eqref{damping} and~\eqref{communication}.
So we can set $\gamma_- = h_- = 0$ for the whole range of temperatures.
In order to model the absence of the Higgs phase at temperatures down to $130$~GeV
we put $\langle \Phi (T) \rangle = 0$.
We consider the NH case and two different values of the common mass, $1$~GeV
and $10$~GeV. The phases are fixed to the values~\eqref{NHphases0}.
We present the results in figure~\ref{diff}.
In order to see how accounting for the charge neutrality of plasma modifies the 
results we replace the susceptibility matrix in~\eqref{susceptibilities} by a diagonal
one. In figure~\ref{sus_sph} we compare results with and without susceptibilities.
One can see that the effect is quite sizeable. In the same figure we demonstrate
the results with and without careful treatment of sphalerons.

Inspecting figures~\ref{diff} and~\ref{sus_sph} one can arrive at the following conclusions.
\begin{itemize}
	\item Fermion number violating rates.\\
	Accounting for the fermion number violation increases the $Y_B$.
	See figure~\ref{diff}, green dashed lines.
	
	\item Broken phase.\\
	Equations without the fermion number violation solved under the assumption that the Higgs vacuum expectation value is zero at all temperatures
	above the $T_{sph}$ lead to larger amount of the $Y_B$ for heavy HNLs
	at large $|\imw|$.
	See figure~\ref{diff}, blue dotted lines.

	\item Neutrality of plasma.\\
	Accounting for the neutrality of plasma by means of susceptibilities
	is important. The effect is stronger for lighter HNLs. See figure~\ref{sus_sph}.

	\item Freeze-out of sphalerons.\\
	The boundary of the allowed region in figure~\ref{sus_sph} is not sensitive
	to the method of calculation of the BAU from the lepton asymmetry.
	In fact, if one is interested in the upper bounds on the 
	mixings (large $|\imw|$) the instantaneous freeze-out of sphalerons can be assumed.
	See figure~\ref{sus_sph}.

\end{itemize}


\section{Study of the parameter space} 
\label{sec:scan_of_the_parameter_spaces}
In this section, we describe how the study of the parameter space of the model have 
been performed. Our strategy is a direct sampling of the 
parameters defining the theory. In  subsection~\ref{sub:bounds_on_the_total_mixings}
we fix specific values of the phases $\delta, \;\eta, \;\rew$
which maximize the generated asymmetry. 
In subsection~\ref{sub:bounds_on_the_individual_mixings} we sample the 
whole $6$ dimensional parameter space.

\subsection{Total mixing} 
\label{sub:bounds_on_the_total_mixings}


In order to set the bound on the value of the total mixing~\eqref{U2}
we need to find, for each value of the common mass, the largest
value of $|U|^2$.

Since the  value of $|U|^2$ for a given mass depends only on $\imw$,
one can marginalize over phases $\delta, \;\eta, \; \rew$ and mass 
difference $\Delta M$
and solve an optimization problem for $\imw$.
The optimization problem consists of 
maximizing (or minimizing for negative values) 
$\imw$ subject to $Y_B \simeq Y_B^{obs}$. 

Several comments are in order.
The value $Y_B$ can be both positive or negative. If it is possible to
obtain some value $Y_B^1$ for some $X_\omega$ and mass 
difference, then it is also  possible to obtain $-Y_B^1$
for the same $X_\omega$ and $\Delta M$ provided that the
 phase parameters in the model can vary freely. 
 In what follows we would always take the absolute value 
$|Y_B|$ of the computed BAU.

Next, it is important to clarify what does  $Y_B \simeq Y_B^{obs}$ actually mean.
The kinetic equations that we solve contain an inherent error stemming from the  assumption of
equilibrium momentum dependence of density matrices.
In order to account for this theoretical uncertainty we impose 
the following condition $Y_B^{obs}/2 < |Y_B| < 2 Y_B^{obs}$.

In practice, it is easier to maximize $|Y_B|$
for given values of $\imw$ and $M$. If the maximal value 
of $|Y_B|$ exceeds e.g. $Y_B^{obs}$, then it is also possible to generate 
a smaller value of asymmetry. 
One can iterate this procedure on a grid in  $\imw$ and $M$ space.
Then by interpolating 
$|Y_B|$ as a function of $\imw$ for a given $M$ and finding roots
 of the equation
$|Y_B| = \kappa Y_B^{obs}$, $\kappa = 0.5, 1, 2$, one can find the upper and lower bounds on $|U|^2$. The case of $\kappa = 2$ corresponds to the conservative
assumption that the averaging procedure amounts to a twice larger asymmetry compared
to the accurate treatment. 
Authors of ref.~\cite{Ghiglieri:2017csp} have solved the full system of equations
for several parameter points. Their results indicate that the averaged equations
rather tend to underestimate the value of BAU. This case is indicated by
$\kappa = 0.5$.

Maximizing $|Y_B|$ with respect to $\Delta M, \rew, \delta, \eta$
is a resource demanding task. It can be significantly simplified in the strong wash-out
regime, i.e. for large values $| \imw |$.
In this regime the value of the total asymmetry 
strongly depends on 
the difference in the damping rates of active neutrinos. 
For a given mass, $X_\omega$
and mass difference the damping rates are controlled by Dirac and Majorana phases together with the real part of $\omega$.
Note that a set of phases that maximizes the difference among
these damping rates (and, the total lepton asymmetry consequently) also 
minimizes (maximizes in the case of IH) $|U_e|^2$.
We have used the following values\footnote{The value $\delta = 0$
for the IH case is incompatible with the recent $3 \sigma$ bounds 
of the NuFit 3.2 analysis. However, setting $\delta= 354^\circ$ doesn't
change the results presented here.}
\begin{subequations}
\begin{align}
	\mbox{NH:} &\quad \delta = \pi , \quad \eta = 3 \pi/2,\quad \rew = \pi/4,
    \label{NHphases}\\
	\mbox{IH:} &\quad \delta =  0,\quad \, \eta = \pi/2,\quad \;\, \rew = \pi/4.
    \label{IHphases}
\end{align}
\label{fixed_phases}\end{subequations}
These choices of phases maximize one of the individual mixings $U_\alpha$ 
(see Appendix~\ref{sec:mixing_angles_of_hnls_and_active_neutrinos}). For the NH case the phases 
\eqref{NHphases} maximize $U_\mu$, while for the IH case the phases~\eqref{IHphases}
maximize $U_e$.
Since the phases are fixed, we need to find only the value of $\Delta M$ which maximizes
BAU at each point of the $M-\imw $ grid. 
The upper bounds in figure~\ref{U2_bounds} were obtained using the method described above.
Let us stress that the same upper bounds can be obtained by the random sampling described 
in the next subsection. We have checked that two methods agree with each other.
The lower bounds on the $|U|^2$ obtained with the 
fixed phases~\eqref{fixed_phases} are not optimal, since the asymmetry is generated in the oscillatory regime. Therefore, the lower bounds in figure~\ref{U2_bounds} are obtained by the direct sampling.


\subsection{Individual mixings} 
\label{sub:bounds_on_the_individual_mixings}

The mixings $|U_\alpha|^2$ depend on $\delta$ and $\eta$ through the 
elements of the PMNS matrix  entering eqs.~\eqref{Cpm_nh} or~\eqref{Cpm_ih}.
Therefore it is no longer possible to solve a simple optimization problem, as was the case for 
$|U|^2$. However, our numerical routines solve kinetic equations for different values of parameters 
very efficiently. This allows us to perform a scan of the full parameter space.

The parameter space is sampled as follows.
As was already mentioned, we are restricted 
to the discrete grid in the common mass $M$ in the interval $[0.1, 10.0]$ GeV.
 The rest of parameters we sample randomly, so that 
$\log_{10} \Delta M, \imw, \rew, \delta, \eta$ are distributed uniformly
in the intervals specified in table~\ref{table_parameters}.\footnote{Note that for the
Dirac phase we have actually used the $3 \sigma$ interval from the NuFit 3.2 analysis. Namely,
$\delta \in [144^\circ, 374^\circ]$ in the NH case, and 
$\delta \in [192^\circ, 354^\circ]$ in the IH case.
}
Note that according to eq.~\eqref{U2A}, the uniform distribution in $\imw$ 
approximately coincides to a uniform distribution 
of $|U_\alpha|^2$ in log-scale. However, in order to obtain the upper bounds more accurately, we also perform a flat sampling in the $X_\omega$.
After computing the value of BAU for each point we select the points according to the  
criterion $|Y_B| > Y_B^{obs}$. 

In order to plot figures \ref{UaUb_NH} and \ref{UaUb_IH} we have generated \num{2800000} points for each hierarchy type and selected only those points for which $|Y_B| > Y_B^{obs}$.
We also have utilized these datasets to obtain the lower bounds
and to cross check the upper bounds on the total mixing $|U|^2$.


\section{Comparison with other works} 
\label{sec:comparison_with_other_works}

\begin{figure}[htb!]
\centerline{
    \includegraphics[width = 0.5\textwidth]{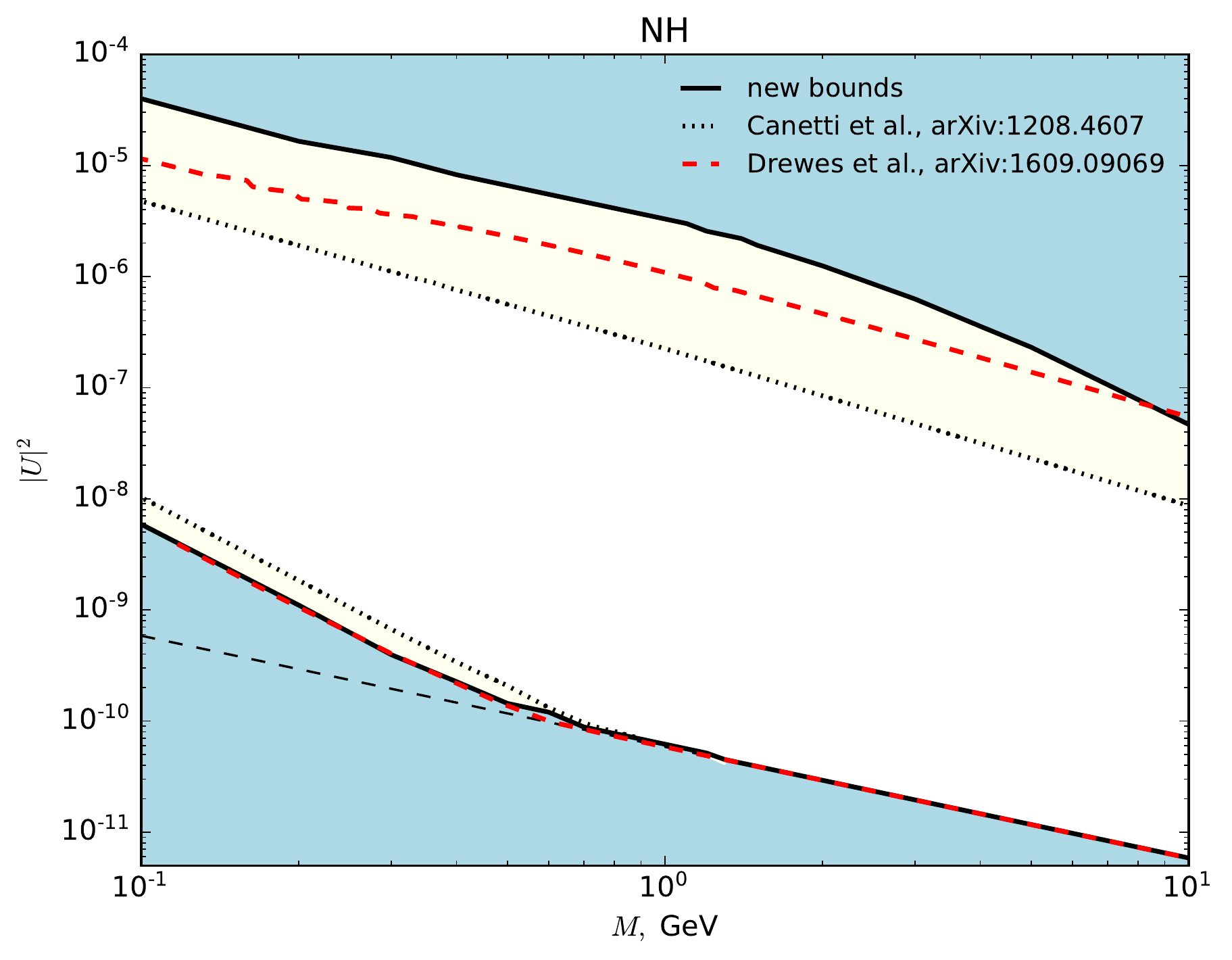}
    \includegraphics[width = 0.5\textwidth]{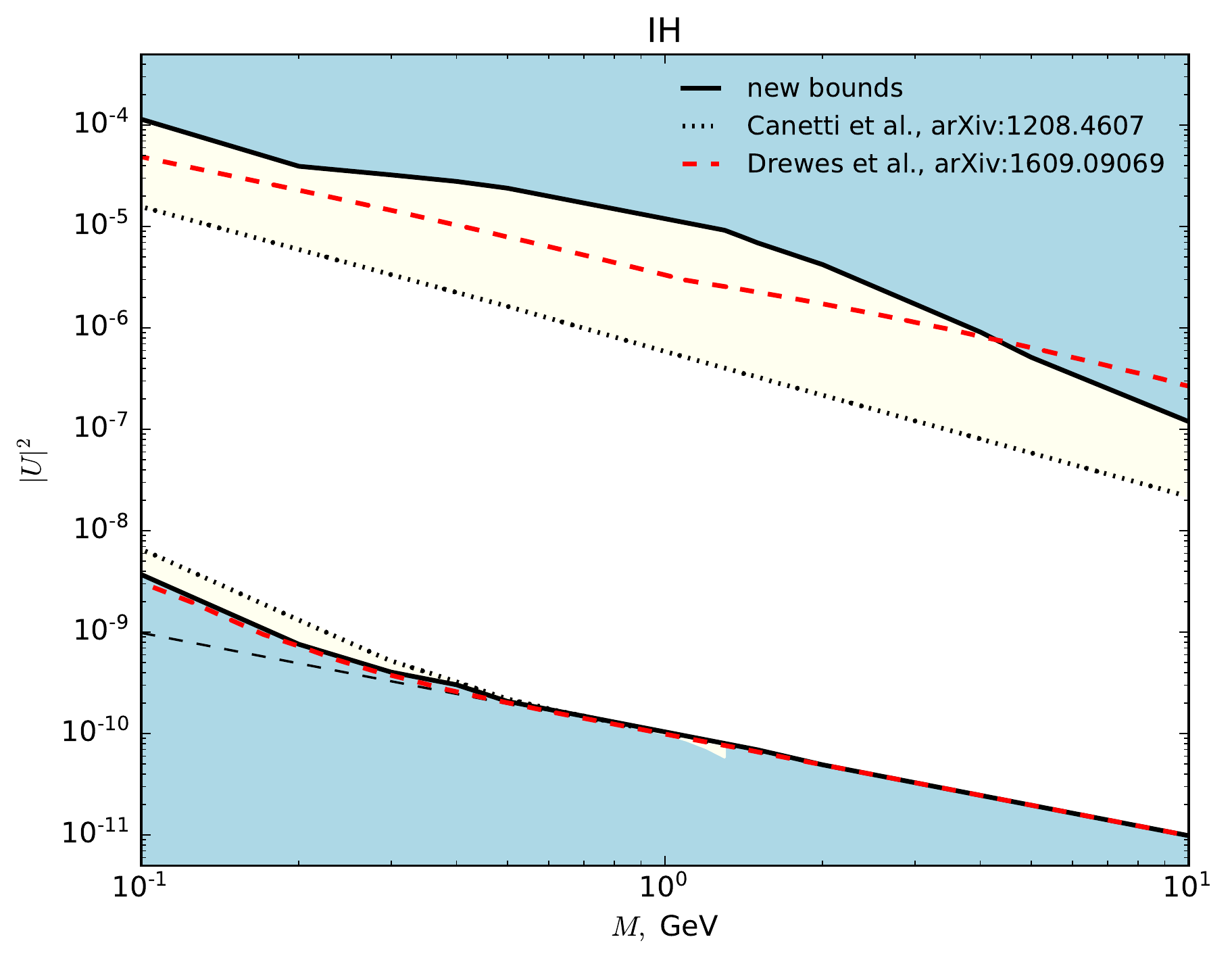}
}
\caption{\label{fig:comparison}Comparison of the bounds from three different works.
Our lower bounds (black solid lines) are obtained from the  parameter sampling,
whereas upper bounds are obtained for fixed phases.
}
\end{figure}
Baryogenesis in the $\nu$MSM has attracted a lot of attention of the 
community in recent  years.
The first scan of the parameter space was performed in 
refs~\cite{Canetti:2012vf, Canetti:2012kh}. 
Authors of ref.~\cite{Shuve:2014zua} have accounted for the
neutrality of the electroweak plasma which leads to 
$\mathcal{O}(1)$ corrections to the final asymmetry.

More recently, scans of the parameter space were performed by
 two groups, see refs.~\cite{Drewes:2016gmt,Hernandez:2016kel}.
The role of fermion number violating processes was clarified in
refs~\cite{Hambye:2016sby, Eijima:2017anv,Ghiglieri:2017gjz,Hambye:2017elz}.
Implications of a non-instantaneous freeze-out of sphalerons
were addressed in refs.~\cite{Eijima:2017cxr,Ghiglieri:2017csp}.

In what follows we list corresponding works.
\begin{description}
	\item \textbf{L. Canetti, M. Drewes, T. Frossard, and M. Shaposhnikov
	\cite{Canetti:2012vf, Canetti:2012kh}}\\
	The first detailed study of the parameter space.
	Only the symmetric phase has been considered.
	Asymmetries in the leptonic sector were described by means
	of the chemical potentials, i.e. neutrality of the plasma has not been 
	accounted for.
	The rates were underestimated by a factor of two 
	(see table~\ref{table_rates} below).
	In the scan of the parameter space the values of phases
	were fixed to non-optimal values.
	As a result, the allowed region of the parameter space is much smaller compared
	to what we have obtained in this work.

	\item \textbf{M. Drewes, B. Garbrecht, D. Gueter, and J. Klaric
		\cite{Drewes:2016gmt,Drewes:2016jae}}\\
	In ref.~\cite{Drewes:2016gmt} only the symmetric phase has been considered.
	The kinetic equations were generalized to the broken phase in 
	ref.~\cite{Drewes:2016jae}. The rescaling of the parameters that simplified
	 computations has been suggested in ref.~\cite{Drewes:2016gmt}.
	The relation between leptonic chemical potentials and number densities
	accounting for the neutrality of the plasma has been implemented.
	This relation is analogous to eq~\eqref{susceptibilities}, however
	it is valid only at large temperatures. In high-temperature limit 
	this relation agrees with eq~\eqref{susceptibilities}.

 	Note the persistent disagreement between the damping rate 
 	of the active neutrinos in ref.~\cite{Drewes:2016gmt} and in our work
 	(see discussion below).



	\item 
	\textbf{P. Hern\'{a}ndez, M. Kekic, J. L\'{o}pez-Pav\'{o}n, J. Racker and J. Salvado
		\cite{Hernandez:2016kel}}\\
	Only the symmetric phase has been considered.
	The neutrality of the plasma has been accounted for, however, apparently,
	the susceptibilities disagree with those in ref.~\cite{Drewes:2016gmt} 
	and with ours at high-temperature limit.\footnote{
It is important to clarify that the matrix  $C_{\alpha \beta}$, entering eq. (2.20) from 
ref.~\cite{Hernandez:2016kel} agrees with  our matrix  $\omega_{\alpha \beta}$ from eq.~\eqref{susceptibilities} provided that in ref.~\cite{Hernandez:2016kel} the symbol $\mu_\alpha$
denotes the chemical potential to left-handed leptons. Note that our $\mu_\alpha$ are the chemical potentials to all leptons of flavour $\alpha$. Therefore, 
$\mu_\alpha^{(ref.~[33])} = \mu_\alpha^{(our)} - \mu_Y/2$,
where $\mu_Y$ is the chemical potential to the hypercharge. We thank Jacopo Ghiglieri
for pointing this out. 
However, once the chemical potentials are eliminated from the kinetic equations by means 
of eqs.~\eqref{susceptibilities} and (2.20) from ref.~\cite{Hernandez:2016kel},
the equations are actually different. Namely, in the r.h.s. of the equations the terms proportional to 
$\sum_\beta \omega_{\alpha, \beta} Y_{\Delta_\beta}$ will appear. The matrices multiplying 
$Y_{\Delta_\beta}$ are different in ref.~\cite{Hernandez:2016kel} and in this work.
  }

	The approach to the study of the parameter space is different from
	what we use in this work. 
	The parameter space has been sampled by means of the Markov Chain Monte
	Carlo (MCMC) with certain priors. The cosmologically allowed regions of the parameter space
	were presented as contours $90\%$ of all generated points.
  This method resulted in regions which are much smaller compared to what we have
	obtained in this work.

	\item \textbf{S. Antusch, E. Cazzato, M. Drewes, O. Fischer, B. Garbrecht, D. Gueter et al
		\cite{Antusch:2017pkq}}\\
	The scan of the parameter space of heavy HNLs ($M>5$~GeV).
	Fermion number violating processes have been accounted for in the symmetric phase.
	The parameter space has been sampled by means of the Markov Chain Monte
	Carlo (MCMC). However, the selection criteria is different 
	from~\cite{Hernandez:2016kel}. Namely, the models leading to
	$Y_B^{obs}- 5 \sigma_{Y_B^{obs}} < |Y_B| < Y_B^{obs} + 5 \sigma_{Y_B^{obs}} $
	were selected.
	This approximately corresponds to $0.68\cdot Y_B^{obs} < |Y_B| < 1.32\cdot Y_B^{obs}$.
	Let us emphasize that the uncertainties in the value of $|Y_B|$
	are theoretical, whereas the experimental  uncertainty, characterized by
	$\sigma_{Y_B^{obs}}$ is much smaller. This is the reason why throughout this
	work we consider a larger interval for $|Y_B|$.

	\item \textbf{J. Ghiglieri and M. Laine~\cite{Ghiglieri:2016xye,Ghiglieri:2017gjz,Ghiglieri:2017csp,Ghiglieri:2018wbs}}\\
	There were no scans of the parameter space. 
	However, a thorough derivation of all rates has been performed. 
	The susceptibilities have been calculated accounting for 
	the non-zero masses of the fermions in ref.~\cite{Ghiglieri:2016xye}.
	The full non-averaged system has been solved for several benchmark points
	in ref.~\cite{Ghiglieri:2017csp}. 

	After the ArXiv version of this paper had been released,
	a new study~\cite{Ghiglieri:2018wbs} appeared. This work contains the
	most up-to-date determination of both fermion number conserving and violating
	rates in the whole temperature region relevant for baryogenesis. It was pointed out in ref.~\cite{Ghiglieri:2018wbs} that the $2k \Gamma_k$ part of the
	$\gamma_{\nu (+)}$ was missing in the ArXiv version of the
	present paper. In the current version of the paper we correct this point.
	We have also updated all rates entering the kinetic equations using the 
	results of ref.~\cite{Ghiglieri:2018wbs}.

   \item \textbf{T. Hambye and D. Teresi~\cite{Hambye:2016sby,Hambye:2017elz}}\\
   There were no scans of the parameter space.
   A role of fermion number violating Higgs decays has been discussed.
   The considerations of ref.~\cite{Hambye:2017elz}
   are limited to the Higgs decays and inverse decays.
   The rate of the fermion number conserving processes
   has been underestimated compared to the  ones including
   $2 \leftrightarrow 2$ scatterings. This can be seen, e.g. from figure 4 of ref.~\cite{Ghiglieri:2017gjz}. Therefore, a direct comparison between our study and ref.~\cite{Hambye:2017elz} is  not straightforward.

\end{description}

It is important to note that the generic structure of kinetic equations is 
the same in all studies of the low-scale leptogenesis. Therefore it is 
possible to compare the rates in the kinetic equations 
independently of their derivation.
In order to be able to compare 
refs~\cite{Canetti:2012kh,Drewes:2016gmt,Hernandez:2016kel,Ghiglieri:2017csp}, we compute the corresponding rates at temperature $T_{ref} = 10^3$~GeV.
At this temperature the rates are dominated by lepton number conserving processes.

The production rate of HNLs, 
the communication term of HNLs, the damping term of the lepton asymmetries 
and their communication term
can be described as
\begin{align}
\overline{\Gamma}_N/h^2 = C_1\cdot T, \\
\overline{\tilde{\Gamma}}_N/h^2 = C_2\cdot T\\
\overline{\Gamma}_\nu/h^2 = C_3 \cdot T\\
\overline{\tilde{\Gamma}}_\nu/h^2 = C_4\cdot T
\end{align}
where $h^2$ is a symbolic representation of an appropriate product of Yukawa 
coupling constants for each term.
The values of the coefficients $C_i$ in considered works are summarized in table~\ref{table_rates}. Note that since authors of ref.~\cite{Ghiglieri:2017csp} treat the
momentum dependence exactly in their numerical computations, we cannot compare their rates, however the hierarchy among the rates and their values at $k = 3 T$ are the same as ours.
\begin{table}[htb!]
	\begin{center}
		\begin{tabular}{| c || c | c | c | c |}
			\hline
			Article & $C_1$ & $C_2$ & $C_3$ & $C_4$ \\ \hline
			This work & 0.0097 & 0.0086 & 0.0086 & 0.0097 \\ \hline
			L. Canetti et al.~\cite{Canetti:2012kh} 
			& 0.005 & 0.005 &0.005 &  0.005 \\ \hline
			M. Drewes et al.~\cite{Drewes:2016gmt} 
			& 0.012 & 0.012 & 0.006 & 0.006 \\ \hline
			P. Hern\'{a}ndez et al.~\cite{Hernandez:2016kel} 
			& 0.0118 & 0.0069 & 0.0076 & 0.0130 \\ \hline
		\end{tabular}
	\end{center}
	\caption{\label{table_rates} The coefficients of the rates in considered works.}
\end{table}
Leaving aside ref.~\cite{Canetti:2012kh}, one can see that the values of the coefficient
$C_1$ do agree with a reasonable precision. However, the values of the other coefficients 
differ roughly by a factor of two from work to work.
In order to understand this difference, we numerically solve our kinetic equations
with the rates multiplied by a constant coefficients $\kappa_a$ $(a=1,2,3,4)$ as follows.
\begin{subequations}
\begin{align}
\overline{\Gamma}_N &\rightarrow \kappa_1 \overline{\Gamma}_N, \\
\overline{\tilde{\Gamma}}_N &\rightarrow \kappa_2 \overline{\tilde{\Gamma}}_N, \\
\overline{\Gamma}_\nu &\rightarrow \kappa_3 \overline{\Gamma}_\nu, \\
\overline{\tilde{\Gamma}}_\nu &\rightarrow \kappa_4 \overline{\tilde{\Gamma}}_\nu.
\end{align}\label{modification}\end{subequations}
Four different cases are considered.
\begin{description}
	\item[case 1] : $\kappa_1 = \kappa_2 = \kappa_3 = \kappa_4 = 1$ for our equations (red lines in the following plot).
	\item[case 2] : $\kappa_1 = \kappa_2 = \kappa_3 = \kappa_4 = \frac{1}{2}$ for ref~\cite{Canetti:2012kh} by L. Canetti et al. (magenta lines).
	\item[case 3] : $\kappa_1 = \kappa_2 = 1$ and $\kappa_3 = \kappa_4 = \frac{1}{2}$ for ref~\cite{Drewes:2016gmt} M. Drewes et al. (blue lines).\footnote{
After the preprint of this paper had been released, we received a comment from the authors
of refs.~\cite{Drewes:2016gmt,Drewes:2016jae}. They found a missing
 factor of two in their calculations.
Once this factor is corrected, the relative sizes of the coefficients $C_i$ in the corresponding row of table~\ref{table_rates}
will approximately agree with these of ref~\cite{Hernandez:2016kel}.
Namely, the \textbf{case 4 } 
will be realized.
  }
	\item[case 4] : $\kappa_1 = \kappa_4 = 1$ and $\kappa_2 = \kappa_3 = \frac{1}{2}$ for ref~\cite{Hernandez:2016kel} by P. Hernandez et al. (green lines).
\end{description}
For the cases 2, 3, and 4 the  values above do not reproduce the kinetic equations in each works exactly, but allow us to understand the qualitative behaviour in each case. 

\begin{figure}[!bth]
 \centerline{
  \includegraphics[clip, width=0.5\textwidth]{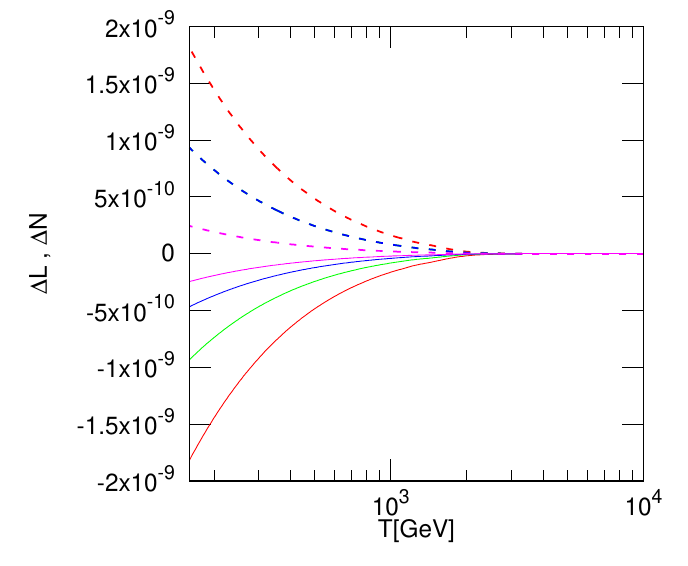}\\
  \includegraphics[clip, width=0.5\textwidth]{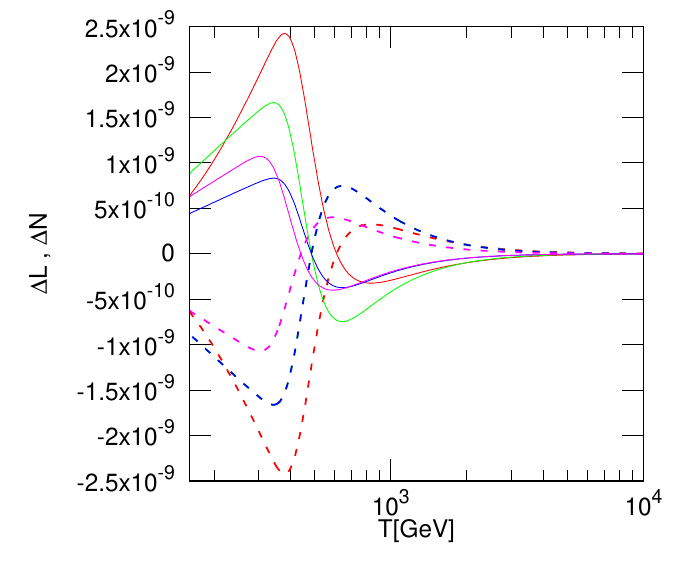}
}
 \caption{Time-evolution of asymmetris; solid lines are sum of asymmetries in 
 the left-handed lepton sector and dashed lines are that of the HNL sector. 
 Note that blue and green dashed lines are overlapped. 
 The common mass $M=1$~GeV, phases are fixed to the values~\eqref{fixed_phases}.
 \emph{Left panel:} $\Delta M = 10^{-7} \, \text{GeV}$ and $\text{Im} \, \omega = 1$. 
 \emph{Right panel:} $\Delta M = 10^{-7} \, \text{GeV}$ and $\text{Im} \, \omega =5$.
 }
 \label{fig:evo_Iw1}
\end{figure}

We demonstrate the time-evolution of asymmetries up to 
$T=160 \, \text{GeV}$ in figure~\ref{fig:evo_Iw1}. 
The qualitative picture of figure~\ref{fig:evo_Iw1} agrees with the results 
presented in figure~\ref{fig:comparison}.

There is also an important comment regarding studies of the parameter space.
In fact, each point in this space defines a \emph{theory}. 
It is not clear at all what could be a prior probability in the space
of theories. The problem is not entirely philosophical.
This can be most easily seen comparing the first columns of sub-plots in figures 4 and 5 from
ref.~\cite{Hernandez:2016kel} with the diagonal sub-plots in
our figures~\ref{UaUb_IH} and~\ref{UaUb_NH}.
Our allowed regions of the parameters space a much larger than
the contours shown in ref.~\cite{Hernandez:2016kel}.
The reason for this difference is that the
study of ref.~\cite{Hernandez:2016kel} relied on a Bayesian analysis
of the Markov Chain Monte Carlo (MCMC). This analysis assumes certain 
prior probabilities in the space of theories and depends strongly on the chosen 
priors~\cite{Hernandez:2016kel}.
 We advocate the point of view that each parameter point leading to the correct
values of the observables (such as neutrino mixing angles and the value of the BAU) 
should be accounted for.


\section{Conclusions and outlook} 
\label{sec:conclusions_and_outlook}
In this work we have performed the thorough study of the parameter space
of baryogenesis in the $\nu$MSM. All important effects have 
been accounted for in our kinetic equations.
Our study improves that of previous works in several respects.

\emph{(i)} The rates entering kinetic equations are calculated from the 
parameters of the theory.
In the symmetric phase, 
as one can see from the table~\ref{table_rates}, in ref.~\cite{Canetti:2012kh}
the values of the rates were consistently underestimated.
Moreover, apart from a factor of two difference in the damping rates,
there is an agreement among all studies. Note also that 
all considered rates are practically the same in our work and 
in ref.~\cite{Ghiglieri:2017csp}.

\emph{(ii)}
In the broken phase the effects of the fermion number violation were 
systematically taken into account for the first time.
These effects are important for the 
baryogenesis even though
the temperature interval between the electroweak crossover and the sphaleron 
freeze-out is rather small. 

\emph{(iii)} 
We have accurately accounted for the sphaleron freeze-out utilizing the
`improved approach' of ref.~\cite{Eijima:2017cxr}.

\emph{(iv)}
Last but not the least improvement is related to the performance of the ODE solver
which was used to solve the kinetic equations numerically. Impressive
increase of efficiency of the numerical routine allowed us to perform 
a comprehensive sampling of the parameter space.

Our main results are upper and lower bounds of the region
where successful baryogenesis in the $\nu$MSM is possible. We list them and stress significant points.
\begin{itemize}
	\item Bounds in the $|U|^2 - M$ plane, figure~\ref{U2_bounds}.
	The allowed region is significantly
	larger for light HNLs compared to the previous studies. Let us emphasize that the position of the
	upper bound is actually important for the direct detection. Even though this 
	region seems to be the easiest for the direct detection owing to the most 
	efficient production of HNLs, it might be actually not the case, because
	the life time of HNLs is short. HNLs can decay before they reach the detector.
	See the line of the SHiP experiment in figure~\ref{U2_bounds}.
	Also, it might be interesting to update
	the study of the neutrinoless double beta decay
	in the $\nu$MSM, refs.~\cite{Drewes:2016lqo,Hernandez:2016kel,Asaka:2016zib}.

	\item Bounds on individual mixings $|U_\alpha|\cdot|U_\beta|$ as functions of $M$.
	Note that we present also the off-diagonal elements. These are important
	for thorough simulations of the experimental sensitivity.
	\item The dataset of different choices of the parameters of the $\nu$MSM. 
	This dataset can be used to compare our approach with other groups.
	As we have already stressed, we use momentum averaged kinetic equations.
	Computation of the BAU in the full system is highly non-trivial and a scan of the parameter space is very demanding. Therefore our parameter sets 
	can be used as benchmark points to test different regimes of the BAU production with the accurate non-averaged equations.
	Models from the dataset could also be used by experimental collaborations
	for Monte Carlo simulations.
\end{itemize}


\acknowledgments
We are grateful to Jacopo Ghiglieri and Mikko Laine for helpful discussions 
and for sharing the numerical data on the direct rates from ref.~\cite{Ghiglieri:2018wbs}.
We thank Alexey Boyarsky, Jacopo Ghiglieri, Mikko Laine, and Oleg Ruchayskiy for helpful comments 
on the paper. We thank Juraj Claric for useful discussions and for sharing the 
data points from ref.~\cite{Drewes:2016jae}. We also thank  
Jacobo L\'{o}pez-Pav\'{o}n for discussions related to the plasma neutrality and the comparison of
kinetic equations.
This work was supported by the ERC-AdG-2015 grant 694896. The work of M.S. and I.T. was supported partially by the Swiss National Science Foundation.

\appendix

\section{Mixings of HNLs and active neutrinos} 
\label{sec:mixing_angles_of_hnls_and_active_neutrinos}

In this appendix we collect the formulae for the mixings of HNLs and active neutrinos.
We considered the two-HNL case here.
All formulae presented here are obtained for the normal hierarchy (NH)
of the neutrino masses. The case of the inverted hierarchy (IH) can be obtained
by the following replacement
\begin{equation}
	\text{NH}\rightarrow \text{IH}:\quad m_2 \rightarrow m_1,\; m_3 \rightarrow m_2,\;
	U^{PMNS}_{\alpha 2} \rightarrow U^{PMNS}_{\alpha 1},\;	U^{PMNS}_{\alpha 3} \rightarrow U^{PMNS}_{\alpha 2}
	\label{NHtoIH}
\end{equation}
So, for example, $m_2+m_3$ becomes $m_1 + m_2$ in the IH case.

The Yukawa coupling constants entering the Lagrangian~\eqref{Lagr}
can be decomposed using the Casas-Ibarra parametrization~\eqref{CasasIbarra}.
Formula~\eqref{CasasIbarra} can be rewritten as
\begin{align}
F_{\alpha 1} &= \frac{\sqrt{M_{1}}}{2 \langle \Phi (0) \rangle} \left[ C_{\alpha}^{+} \tilde{X}_{\omega} + C_{\alpha}^{-} \tilde{X}_{\omega}^{-1} \right], \\
F_{\alpha 2} &= i \frac{\sqrt{M_{2}}}{2 \langle \Phi (0) \rangle} \left[ C_{\alpha}^{+} \tilde{X}_{\omega} - C_{\alpha}^{-} \tilde{X}_{\omega}^{-1} \right],
\end{align} 
where
\begin{align}
 M_{1} &= M - \Delta M, \\
 M_{2} &= M + \Delta M, \\
\tilde{X}_{\omega} &= X_{\omega} e^{- i \text{Re}\, \omega}, \\
C_{\alpha}^{+} &= i U^{PMNS}_{\alpha 2} \sqrt{m_{2}} + 
\xi U^{PMNS}_{\alpha 3} \sqrt{m_{3}}, \label{cp}\\
C_{\alpha}^{-} &= i U^{PMNS}_{\alpha 2} \sqrt{m_{2}} - 
\xi U^{PMNS}_{\alpha 3} \sqrt{m_{3}}, \label{cm}
\end{align}
the Higgs vev at zero temperature is $\langle \Phi (0) \rangle=174.1$ GeV. 
In the case of two HNLs the PMNS matrix contains two phases:
\begin{equation}
U^{PMNS} =  \left(
\begin{array}{ccc}
 c_{12} c_{13} & s_{12} c_{13} e^{i \eta } & s_{13} e^{-i \delta }\\
 -s_{12} c_{23} - c_{12} s_{13} s_{23} e^{i \delta } & \left(c_{12} c_{23} - s_{12} s_{13} s_{23} e^{i \delta } \right) e^{i \eta } & c_{13} s_{23} \\
 s_{12} s_{23} - c_{12} s_{13} c_{23} e^{i \delta } & - \left(c_{12} s_{23} + s_{12} s_{13} c_{23} e^{i \delta } \right) e^{i \eta } & c_{13} c_{23} \\
\end{array}
\right).
\end{equation}

Up to the leading order of the seesaw mechanism the mixing elements of HNLs are 
\begin{align}
 \Theta_{\alpha 1} &= \frac{\langle \Phi (0) \rangle F_{\alpha 1}} {M_{1}} = \frac{1}{2 \sqrt{M_{1}}} \left[ C_{\alpha}^{+} \tilde{X}_{\omega} + C_{\alpha}^{-} \tilde{X}_{\omega}^{-1} \right], \\
  \Theta_{\alpha 2} &= \frac{\langle \Phi (0) \rangle F_{\alpha 2}} {M_{2}} = \frac{i}{2 \sqrt{M_{2}}} \left[ C_{\alpha}^{+} \tilde{X}_{\omega} - C_{\alpha}^{-} \tilde{X}_{\omega}^{-1} \right].
\end{align}

It is possible to show that the flavour components of the mixings are given by
\begin{align}
 |U_{\alpha}|^{2} &= \sum_{I} |\Theta_{\alpha I}|^{2} \\
                           &= \frac{1}{2 (M^{2} - \Delta M^{2})} \left[ M (|C_{\alpha}^{+}|^{2} X_{\omega}^{2} + |C_{\alpha}^{-}|^{2} X_{\omega}^{-2}) + 2 \Delta M \mathrm{Re}\, [C_{\alpha}^{+} (C_{\alpha}^{-})^{\ast} e^{- i 2 \mathrm{Re}\, \omega} ] \right] \label{Ua2}\\
                           &\simeq \frac{1}{2 M} \left[ (|C_{\alpha}^{+}|^{2} X_{\omega}^{2} + |C_{\alpha}^{-}|^{2} X_{\omega}^{-2}) + 2 \frac{\Delta M}{M} \mathrm{Re}\, [C_{\alpha}^{+} (C_{\alpha}^{-})^{\ast} e^{- i 2 \mathrm{Re}\, \omega}] \right] \label{Ua2A} \\
                           &\simeq \frac{1}{2 M} \left[ |C_{\alpha}^{+}|^{2} X_{\omega}^{2} + |C_{\alpha}^{-}|^{2} X_{\omega}^{-2} \right], \label{Ua2AA}
\end{align}
where
\begin{align}
&|C_{\alpha}^{+}|^{2} = |U^{PMNS}_{\alpha 2}|^{2} m_{2} + |U^{PMNS}_{\alpha 3}|^{2} m_{3} - 2 \xi \sqrt{m_{2} m_{3}} \, \mathrm{Im}\, [U^{PMNS}_{\alpha 2} U_{\alpha 3}^{PMNS \ast}], \label{cp2}\\
&|C_{\alpha}^{-}|^{2} = |U^{PMNS}_{\alpha 2}|^{2} m_{2} + |U^{PMNS}_{\alpha 3}|^{2} m_{3} + 2 \xi \sqrt{m_{2} m_{3}} \, \mathrm{Im}\, [U^{PMNS}_{\alpha 2} U_{\alpha 3} ^{PMNS \ast}], \label{cm2}\\
&\mathrm{Re}\, [C_{\alpha}^{+} (C_{\alpha}^{-})^{\ast} e^{- i 2 \mathrm{Re}\, \omega}] = (|U_{\alpha 2}|^{2} m_{2} - |U^{PMNS}_{\alpha 3}|^{2} m_{3}) \cos (2 \mathrm{Re}\, \omega) - \\ & \phantom{\mathrm{Re}\, [C_{\alpha}^{+} (C_{\alpha}^{-})^{\ast} e^{- i 2 \mathrm{Re}\, \omega}] = }
2 \xi \sqrt{m_{2} m_{3}} \, \mathrm{Re}\, [U^{PMNS}_{\alpha 2} U_{\alpha 3}^{PMNS \ast}] \sin (2 \mathrm{Re}\, \omega). \label{cpcm}
\end{align}

Using  the unitarity of the PMNS matrix one can derive the following expression 
for the total mixing
\begin{align}
|U|^{2} &= \sum_{\alpha} |U_{\alpha}|^{2} \\
            &= \frac{1}{2 (M^{2} - \Delta M^{2})}\left[ M (m_{2} + m_{3}) (X_{\omega}^{2} + X_{\omega}^{-2}) + 2 \Delta M (m_{2} - m_{3}) \cos (\text{Re}\, \omega)\right] \label{U2_app} \\
            &\simeq \frac{1}{2 M}\left[ (m_{3} + m_{2}) (X_{\omega}^{2} + X_{\omega}^{-2}) - 2 \frac{\Delta M}{M} (m_{3} - m_{2}) \cos (\text{Re}\, \omega)\right] \label{U2A} \\
            &\simeq \frac{1}{2 M}\left[ (m_{3} + m_{2}) (X_{\omega}^{2} + X_{\omega}^{-2}) \right]. \label{U2AA}
\end{align}

Corresponding formulae in the IH case can be obtained by means of the 
replacement~\eqref{NHtoIH}.


\section{Derivation of the kinetic equations in Higgs phase} 
\label{sec:derivation_of_kinetic_equations}

In this appendix we present the derivation of the kinetic equations in the Higgs phase. We consider only processes where the Higgs phase is substituted by its vev, i.e. only indirect processes (see, e.g. eq.~\eqref{gamma_pm}) which give
the dominant contribution.

The kinetic equations accounting for the fermion number violating processes 
in the Higgs phase were derived in ref.~\cite{Eijima:2017anv}. 
A certain ansatz about the modification of the neutrino energies by the SM plasma
has been made there. 
Here we  extent the method of ref.~\cite{Eijima:2017anv} so that the interactions with the SM particles are consistently accounted for. This consideration is motivated by recent ref.~\cite{Ghiglieri:2018wbs} where an 
extra active neutrino rate, missing in the equations of ref.~\cite{Eijima:2017anv}, is involved. Our derivation here confirms the results of ref.~\cite{Ghiglieri:2018wbs}.

Below we derive the kinetic equations~\eqref{KE_1a}.
First, we overview the idea behind the derivation and then
present the actual calculations.

\subsection{Overview of the procedure} 
\label{sub:overview_of_the_procedure}
\begin{itemize}
  \item The lepton asymmetry in the $\nu$MSM is generated due to interactions of 
  active neutrinos with HNLs and coherent oscillations of the latter.
  Therefore we need to derive the equations describing the evolution of 
  number densities of both active neutrinos, $\rho_{\nu_\alpha}$ and HNLs 
  $\rho_{N_I}$ as well as correlations of HNLs.


  \item We  work in the Heisenberg picture and introduce a time-independent 
  density matrix of the complete system $\boldsymbol{\rho}$. The distribution 
  function of a particle created by an operator $a^\dagger$ is given then by
  $\Tr[a^\dagger a\, \boldsymbol{\rho}]$. 

  \item The time evolution of an operator $\mathcal{O}$
  is governed by the Heisenberg equation
  \begin{equation}
    \frac{d}{dt} \mathcal{O}(t) = i [ H, \mathcal{O}(t) ],
    \label{Heisenberg0}
  \end{equation}
  where $H$ is the  Hamiltonian of the system. We are interested in the operators
  of the type $\mathcal{O} = a^\dagger a$. So we need 
  to derive the Hamiltonian in terms of creation and annihilation operators.

  \item The evolution equations for the operators describing the number densities
  of neutrinos and HNLs and correlations of HNLs involve some new operators.
  We write down the evolution equations for these new operators. 
  These equations, in turn, involve new operators. We keep going 
  until the system of the equations closes (note that this is very
  different from the Bogolyubov-Born-Green-Kirkwood-Yvon hierarchy which should be 
  truncation at some level).

  \item We obtain a set of a large number of first-order ordinary 
  differential equations. Noticing  that  distinct time 
  scales are present in these equations, one can integrate out  fast oscillations
  and obtain a system describing the slow evolution of the quantities of interest 
  (number densities and correlations).

\end{itemize}

\subsection{Lagrangian and Hamiltonian} 
\label{sec:lagrangian_and_hamiltonian}
The Lagrangian in the mass basis~\eqref{Lagr} is useful for a study 
of the phenomenology of the $\nu$MSM. For the derivation of the kinetic equations
it is more convenient to change the basis~\cite{Shaposhnikov:2008pf}
so the Lagrangian reads (we use tilde $\tilde{N}_I$ to indicate the different basis)
\begin{subequations}
\begin{eqnarray}
\mathcal{L}_\text{SM+2RH$\nu$} &=& \mathcal{L}_0 + \Delta \mathcal{L}, \\
\mathcal{L}_0 &=& \mathcal{L}_\text{SM} + \overline{\tilde{N}_I} i \partial_\mu \gamma^\mu \tilde{N}_I - (h_{\alpha 2} \overline{L_\alpha} \tilde{N}_2 \tilde{\Phi} + M \overline{\tilde{N}_2^c} \tilde{N}_3 + h.c.), \\
\Delta \mathcal{L} &=& - h_{\alpha 3} \overline{L_\alpha} \tilde{N}_3 \tilde{\Phi} - \frac{\Delta M}{2} \overline{\tilde{N}_I^c} \tilde{N}_I + h.c.\,,
\end{eqnarray}
\label{Lagrangian_2}\end{subequations}
where $\Delta M$ is the Majorana mass difference so that the mass
matrix of two heavier HNLs is $M_I = diag(M-\Delta M,M+\Delta M)$.
The matrix of Yukawa couplings $h_{\alpha I}$ can be related to the matrix
$F_{\alpha I}$ defined in~\eqref{Lagr} as follows
\begin{eqnarray}
F_{\alpha I} &=& h_{\alpha J} [U_N^\ast]_{JI}, \\
U_N &=& \frac{1}{\sqrt{2}}\begin{pmatrix} -i & 1 \\ i & 1 \end{pmatrix}.
\end{eqnarray}

It is convenient to further rewrite the Lagrangian~\eqref{Lagrangian_2}
by unifying two Majorana fields into one Dirac field $\Psi = N_2^c + N_3$.
After that, the Lagrangian in the Higgs phase reads
\begin{align}
 \mathcal{L} &=  \mathcal{L}_{SM} +   \overline{\Psi} i \partial_\mu \gamma^\mu \Psi
 - M \overline{\Psi}\Psi +  \mathcal{L}_{int},\label{Lagrangian} \\
\mathcal{L}_{int} &=
- \frac{\Delta M}{2} (\overline{\Psi}\Psi^c + \overline{\Psi^c}\Psi) 
 - (h_{\alpha 2} \langle \Phi \rangle \overline{\nu_{L \alpha}} \Psi + h_{\alpha 3} \langle \Phi \rangle \overline{\nu_{L\alpha}} \Psi^c + h.c.),
 \label{Lint}
\end{align}
where $\mathcal{L}_{SM}$ is the SM part,  $M = (M_{3} + M_{2})/2$
and $\Delta M = (M_{3} - M_{2})/2$ are the common mass and Majorana mass difference, 
$\langle \Phi \rangle = \langle \Phi(T) \rangle$ is
the temperature dependent Higgs vacuum expectation value, $\langle \Phi (0) 
\rangle = 174.1$~GeV. 

We treat the mass difference of HNLs and their interactions with 
left-handed neutrinos as small perturbations. It is important that all the SM 
interactions---including these of active neutrinos---occur with much larger rates 
compared to those originating from the~\eqref{Lint}. It is therefore reasonable
to formulate a perturbation theory in small parameters of~\eqref{Lint}. In what follows we
realize this program.

The momentum expansion of the HNLs field is given by
\begin{equation}
    \Psi(x)=\int \frac{d^3\bp}{(2\pi)^3\sqrt{2 E_p}}\sum_{s = \pm} 
     \left(a_s(\bp)\,u(p,s) e^{-ipx} + b_s^\dagger(\bp)\,v(p,s)e^{ipx} \right).
     \label{psi_fileds}
\end{equation}
We orient $\bp$ along the $z$ axis.
choose plane wave solutions $u(p,s)$ and $v(p,s)$
the helicity states of $\Psi$ are $s = \pm$
and the operators $a_s$ and $b_s$
obey the usual anticommutation relations 
\begin{equation}
    \begin{aligned}
\{ a_h(\bp), a_{h'}^\dagger(\bp') \} &= (2 \pi)^3
    \delta_{h h'} \cdot \delta(\bp - \bp'),\\
\{ b_h(\bp), b_{h'}^\dagger(\bp') \} &= (2 \pi)^3
    \delta_{h h'} \cdot \delta(\bp - \bp')
    \end{aligned}
\end{equation}
with all other anticommutators equal to zero.
We also introduce operators $a_{\nu_\alpha}^\dagger(\bp)$,
$a_{\nu_\alpha}(\bp)$, $b_{\nu_\alpha}^\dagger(\bp)$ and 
$b_{\nu_\alpha}(\bp)$ describing the SM neutrinos
and antineutrinos correspondingly, $\alpha = e, \mu, \tau$.
These operators obey analogous anticommutation relations.

In the HNL sector, we assign a positive fermion number to a particle
with a positive helicity 
and to an antiparticle with a negative helicity.
For instance, one HNL is created by $a_+^\dagger(\pm \bp)$ and another one 
is created by $b_-^\dagger(\pm\bp)$,
see table~\ref{operators_particles}.
Attributed in this way,
the fermion number is conserved in the limit ${ M\to 0, \Delta M \to 0 }$.
\begin{table}[htb!]
\begin{center}
  \begin{tabular}{| c | c | c |}
    \hline
   & particles & antiparticles \\ \hline
  HNLs & $a_+^\dagger(\pm \bp), b_-^\dagger(+\bp)$  & 
         $a_-^\dagger(\pm \bp), b_+^\dagger(+\bp)$ \\ \hline 
  neutrinos & $a_{\nu_\alpha}^\dagger(\pm\bp)$ & $b_{\nu_\alpha}^\dagger(\pm\bp)$\\
    \hline
  \end{tabular}
\end{center}
\caption{\label{operators_particles} Creation operators for particles
and antiparticles.}
\end{table}

We will work in the matrix of densities 
formalism inspired by ref.~\cite{Sigl:1992fn}. 
At the end of the day, we want to describe the 
distribution functions of HNLs and their coherent oscillations. 
Therefore we are interested in time evolution of bilinears of the type
$ \mathcal{O} =  a_N^\dagger(\bp, +1) a_N(\bp, +1)$. The time evolution of such operators
is governed by the Heisenberg equation
\begin{equation}
  \frac{d}{dt} \mathcal{O}(t) = i [ H, \mathcal{O}(t) ],
  \label{Heisenberg}
\end{equation}
where $H$ is the total Hamiltonian of the system.
This Hamiltonian can be decomposed as
\begin{equation}
H = H_0+H_{int}+H_{int}^{SM}.
\label{fullH}
\end{equation}
In the last expression, $H_0$ is the Hamiltonian of the free Dirac field $\Psi$ 
and the free Weyl
fields $\nu_{L_\alpha}$; $H_{int}$ describes \emph{quadratic} interactions of the 
HNLs and left-handed neutrinos and $H_{int}^{SM}$ is the Hamiltonian describing 
all SM interactions of $\nu_{L_\alpha} $.

In order to be able to use eq.~\eqref{Heisenberg} we  express
the Hamiltonian~\eqref{fullH} in terms of creation and 
annihilation operators. For the first and the second terms in eq~\eqref{fullH}
it is a tedious but straightforward task. 
The SM part has to be treated separately.

Using~\eqref{psi_fileds} and analogous decomposition for the neutrino fields,
 one can find that the free Hamiltonian $H_0$ is given by
\begin{align}
  H_0 &= \int \frac{d^3\bp}{(2\pi)^3}\sum_{h=\pm1} E_N(\bp)\left( 
  a_N^\dagger(\bp, h)a_N(\bp, h) + b_N^\dagger(\bp, h)b_N(\bp, h)
   \right) +\\
   & \phantom{=} \sum_{\alpha = e, \mu, \tau}  
   \int \frac{d^3\bp}{(2\pi)^3} E_{\nu_\alpha}\left( 
   a_{\nu_\alpha}^\dagger(\bp)a_{\nu_\alpha}(\bp) + b_{\nu_\alpha}^\dagger(\bp)b_{\nu_\alpha}(\bp)
    \right),
    \label{H0}
\end{align}
where $E_N(\bp)$ and  $E_{\nu_\alpha}(\bp)$ are the energies of 
HNLs and active neutrinos.
As the SM model effects are accounted for, the energy of the active neutrino must be
replaced by a temperature dependent dispersion relation, see e.g. 
refs~\cite{Notzold:1987ik,Morales:1999ia}. We will use the same 
symbol $E_{\nu_\alpha}$ for both vacuum and thermal energies of neutrinos.
Note, however, that $E_{\nu_\alpha}$ in medium can deviate significantly
from the vacuum value which is just $|\bp|$ 
(the small neutrino masses can be safely neglected).

The interaction Hamiltonian $H_{int}$ can be further
decomposed into the Majorana and Yukawa parts coming from the first and the second
parenthesis in~\eqref{Lint} respectively
\begin{equation}
  H_{int} = H_{int}^{Majorana} + H_{int}^Y.
\end{equation}
The Majorana part of the interaction Hamiltonian is
\begin{align}
&H_{int}^{Majorana} = \int \frac{d^3\bp}{(2 \pi)^3} \mathcal{H}^M,\label{HintMaj} \\
&\begin{aligned}
\mathcal{H}^M = 
&\frac{ p \Delta M}{E_N} \left( a_-(\bp)    a_-(-\bp) +  a_+(\bp)    
a_+(-\bp) +  b_+^\dagger(\bp)    b_+^\dagger(-\bp) +  
b_-^\dagger(\bp)    b_-^\dagger(-\bp)\right) + \\
&\frac{M \Delta M}{E_N} \left( 
b_+(\bp)    a_-^\dagger(\bp) +  b_-(-\bp)    a_+^\dagger(-\bp) +  
a_+^\dagger(\bp)    b_-(\bp) +  a_-^\dagger(-\bp)    b_+(-\bp)\right) +\\ 
&\frac{p \Delta M}{E_N} \left( b_+(-\bp)    b_+(\bp) +  b_-(-\bp) 
b_-(\bp) +  a_-^\dagger(-\bp)    a_-^\dagger(\bp) +  
a_+^\dagger(-\bp)    a_+^\dagger(\bp)\right) + \\
&\frac{M \Delta M}{E_N} \left( 
a_-(\bp)    b_+^\dagger(\bp) +  a_+(-\bp)    b_-^\dagger(-\bp) +  
b_-^\dagger(\bp)    a_+(\bp) +  b_+^\dagger(-\bp)    a_-(-\bp)\right),
\end{aligned}\label{HintMaj_operators}
\end{align}
where $p\equiv |\bp|$.

The part of the interaction Hamiltonian coming
from the Yukawa interactions reads
\begin{align}
&H_{int}^Y = \sum_{\alpha = e, \mu, \tau} \int \frac{d^3\bp}{(2\pi)^3} 
\mathcal{H}^Y\label{Hint}\\
&\begin{aligned}
\mathcal{H}^Y = 
&h_{\alpha 2}^* K(\bp) \left(- b_-(\bp)   b_{\nu_\alpha}^\dagger(\bp) -  
b_-(-\bp)   b_{\nu_\alpha}^\dagger(-\bp) +  a_-^\dagger(\bp)   
a_{\nu_\alpha}(\bp) +  a_-^\dagger(-\bp)   a_{\nu_\alpha}(-\bp)\right) +\\
&h_{\alpha 2} K(\bp) \left(- b_{\nu_\alpha}(\bp)   b_-^\dagger(\bp) -  
b_{\nu_\alpha}(-\bp)   b_-^\dagger(-\bp) +  a_{\nu_\alpha}^\dagger(\bp)   
a_-(\bp) +  a_{\nu_\alpha}^\dagger(-\bp)   a_-(-\bp)\right)-\\
& h_{\alpha 3}^* K(\bp) \left( a_+(\bp)   b_{\nu_\alpha}^\dagger(\bp) 
-  a_+(-\bp)   b_{\nu_\alpha}^\dagger(-\bp) +  
b_+^\dagger(\bp)   a_{\nu_\alpha}(\bp) - 
 b_+^\dagger(-\bp)   a_{\nu_\alpha}(-\bp)\right) -\\
&h_{\alpha 3} K(\bp) \left( b_{\nu_\alpha}(\bp)   a_+^\dagger(\bp) -  b_{\nu_\alpha}(-\bp)   a_+^\dagger(-\bp) +  a_{\nu_\alpha}^\dagger(\bp)   b_+(\bp) -  a_{\nu_\alpha}^\dagger(-\bp)   b_+(-\bp)\right)-\\
&h_{\alpha 2}^* \tilde{K}(\bp) \left( b_+(\bp)   a_{\nu_\alpha}(-\bp) +  
b_+(-\bp)   a_{\nu_\alpha}(\bp) -  
a_+^\dagger(\bp)   b_{\nu_\alpha}^\dagger(-\bp) - 
 a_+^\dagger(-\bp)   b_{\nu_\alpha}^\dagger(\bp)\right) -\\
&h_{\alpha 2} \tilde{K}(\bp) \left(- b_{\nu_\alpha}(\bp)   a_+(-\bp) -
 b_{\nu_\alpha}(-\bp)   a_+(\bp) +  a_{\nu_\alpha}^\dagger(\bp)   b_+^\dagger(-\bp) +
 a_{\nu_\alpha}^\dagger(-\bp)   b_+^\dagger(\bp)\right)-\\
&h_{\alpha 3}^* \tilde{K}(\bp) \left(- a_-(\bp)   a_{\nu_\alpha}(-\bp) + 
 a_-(-\bp)   a_{\nu_\alpha}(\bp) - 
  b_-^\dagger(\bp)   b_{\nu_\alpha}^\dagger(-\bp) + 
   b_-^\dagger(-\bp)   b_{\nu_\alpha}^\dagger(\bp)\right) - \\
   &h_{\alpha 3} \tilde{K}(\bp) \left( b_{\nu_\alpha}(\bp)   b_-(-\bp) - 
    b_{\nu_\alpha}(-\bp)   b_-(\bp) + 
     a_{\nu_\alpha}^\dagger(\bp)   a_-^\dagger(-\bp) -  
     a_{\nu_\alpha}^\dagger(-\bp)   a_-^\dagger(\bp)\right),
\end{aligned}
\label{HintY_operators}
\end{align} 
where \begin{equation}
K(\bp) = \frac{\langle \Phi \rangle \sqrt{E_\nu (E_N - p)}}{\sqrt{2 E_N E_\nu}} , \quad
\tilde{K}(\bp) = \frac{\langle \Phi \rangle 
\sqrt{E_\nu (E_N + p)}}{\sqrt{2 E_N E_\nu}}.
\label{F_func}
\end{equation}

The interactions of active neutrinos with the SM plasma are described by the
following Hamiltonian.
\begin{equation}
  H_{int}^{SM} = 
  \int d^3x \,G_F \left(\bar{\nu}_{L \alpha} J + J^\dagger \nu_{L \alpha}\right),
\end{equation}
where  $J$ is the SM current coupled to neutrino. It is important that
$J$ anticommutes with the creation and annihilation operators from 
table~\ref{operators_particles}.

\subsection{Commutators and averaging} 
\label{sub:the_full_set_of_equations}
Now we can use the Hamiltonian~\eqref{fullH} to derive the evolution equations.
However, first we need to identify the operators of interest.

Note that the interaction Hamiltonians 
\eqref{HintMaj} and~\eqref{Hint} contain both positive and negative $\bp$.
The distribution functions of HNLs and active neutrino can be constructed from
the operators 
\begin{equation}
  \begin{aligned}
  &a_+^\dagger(+\bp) a_+(+\bp),\; a_+^\dagger(+\bp) b_-(+\bp),\;
  b_-^\dagger(+\bp) a_+(+\bp),\; b_-^\dagger(+\bp) b_-(+\bp), \;\\
  &a_{\nu_\alpha}^\dagger(+\bp) a_{\nu_\alpha}(+\bp), \;
  \end{aligned}
  \label{slow_pos}
\end{equation}
or from the operators with inverted sign of the spatial momentum
\begin{equation}
  \begin{aligned}
  &a_+^\dagger(-\bp) a_+(-\bp),\; a_+^\dagger(-\bp) b_-(-\bp),\;
  b_-^\dagger(-\bp) a_+(-\bp),\; b_-^\dagger(-\bp) b_-(-\bp), \;\\
  &a_{\nu_\alpha}^\dagger(-\bp) a_{\nu_\alpha}(-\bp). \;
\end{aligned}
\label{slow_neg}
\end{equation}
Operators~\eqref{slow_pos} and~\eqref{slow_neg} will be mixed after 
commuting them with 
the interaction Hamiltonians. 
Nevertheless, this apparent duplication is not a problem. 
Notice that operators~\eqref{slow_neg} can be obtained from~\eqref{slow_pos}
by a parity transformation. Combining operators~\eqref{slow_pos} with those obtained by a parity transformation
one can define
\begin{equation}
  \begin{aligned}
  \mathcal{O}_{11} &=a_+^\dagger(+\bp) a_+(+\bp)+a_+^\dagger(-\bp) a_+(-\bp),\\
  \mathcal{O}_{12} &=a_+^\dagger(+\bp) b_-(+\bp) - a_+^\dagger(-\bp) b_-(-\bp),\\
  \mathcal{O}_{21} &=b_-^\dagger(+\bp) a_+(+\bp) - b_-^\dagger(-\bp) a_+(-\bp),\\
  \mathcal{O}_{22} &=b_-^\dagger(+\bp) b_-(+\bp)+b_-^\dagger(-\bp) b_-(-\bp),\\
  \mathcal{O}_{\nu_\alpha} &=a_{\nu_\alpha}^\dagger(+\bp) a_{\nu_\alpha}(+\bp)+
    a_{\nu_\alpha}^\dagger(-\bp) a_{\nu_\alpha}(-\bp), \; \alpha = e, \mu, \tau,
  \end{aligned}
  \label{slow_comb}
\end{equation}
where the minus signs in~\eqref{slow_comb} appear as
a consequence of the negative intrinsic parity of $b_\pm$ 
in our representation. The operators describing the antiparticles are 
constructed in full analogy with~\eqref{slow_comb} using the definitions in 
table~\ref{operators_particles}.

Now we commute the operators~\eqref{slow_comb} with the full 
Hamiltonian~\eqref{fullH} in order to write down the evolution 
equations~\eqref{Heisenberg}. Let us illustrate the procedure 
considering  the operator
$a_+^\dagger(+\bp) a_+(+\bp)$ only. 
The Heisenberg equation for this operator reads
\begin{equation}
  -i \frac{d}{dt} \left( a_+^\dagger(\bp) a_+(\bp) \right) =
  \left[H, a_+^\dagger(\bp) a_+(\bp)  \right] = h_{e\,2} K(\bp) 
  a_{\nu_e}^\dagger(\bp)a_+^\dagger(\bp) + \ldots
  \label{comm1}
\end{equation}
The commutator contains 16 different operators, most of them are other then
\eqref{slow_pos} and~\eqref{slow_neg}. We have explicitly shown only one 
new operator, let us denote it 
$\mathcal{O}^{fast}_{1} \equiv a_{\nu_e}^\dagger(\bp)a_+^\dagger(\bp)$.
It is important that all terms on the r.h.s. of~\eqref{comm1}
are proportional to small parameters, either Yukawa couplings or
Majorana mass difference.
Now we need
the Heisenberg equation for $\mathcal{O}^{fast}_{1}$, which is
\begin{equation}
  -i \frac{d}{dt}\mathcal{O}^{fast}_{1} =
  \left[H, \mathcal{O}^{fast}_{1}  \right] = (E_{\nu_e} - E_N) 
  \mathcal{O}^{fast}_{1}+ \sum_k C_k \mathcal{O}^{fast}_{k}+\sum_l C_l' \mathcal{O}^{slow}_{l}+
  \sum_l C_m'' \mathcal{O}^{medium}_{m},
  \label{comm2}
\end{equation}
where the first term on r.h.s. arises due to the commutation of $\mathcal{O}^{fast}_1$ with $H_0$, the terms $\mathcal{O}^{fast}_{k}$ and $\mathcal{O}^{slow}_{l}$
come from the commutation with $H_{int}$ and operators $\mathcal{O}^{medium}_{m}$
come from the commutation with $H_{int}^{SM}$.

Note the difference between ~\eqref{comm1} and~\eqref{comm2}:
the time derivative of $a_+^\dagger(\bp) a_+(\bp)$ 
is proportional to the small parameter $h_{e\,2}$, whereas
the time derivative $\mathcal{O}^{fast}_{1}$ is proportional to 
$i (E_{\nu_e} - E_N) \mathcal{O}^{fast}_{1}$. 
In what follows we will call ``slow'' the operators whose time derivatives 
are proportional 
to the small parameters, such as Yukawas. All other operators are ``fast''.
All operators~\eqref{slow_comb} (as well as~\eqref{slow_pos} and
\eqref{slow_neg} separately) do commute with the free 
Hamiltonian~\eqref{H0} and therefore they are of the slow type.
The operators $\mathcal{O}^{fast}_{k}$ present in the commutators of~\eqref{slow_comb} with the full 
Hamiltonian are of the fast type.

We derive equations analogous to \eqref{comm2} for all operators appearing
in the commutators of the operators \eqref{slow_comb} with $H$. 
These equations would form
a closed set if not the SM interactions. Indeed, there are no equations for terms $C_m'' \mathcal{O}^{medium}_{m}$. These terms have the form, e.g.,
 $\bar{u}_\nu(-\bp, h) a_{h}(\bp) J(\-bp)$, where
$J(\bp)$ is the Fourier transform of the medium current $J(x)$.
We now derive the evolution equation for $\mathcal{O}^{medium}_{m}$. It has
the following form.
\begin{equation}
  -i \frac{d}{dt}\mathcal{O}^{medium}_{m} =
  \left[H, \mathcal{O}^{medium}_{m}  \right] = E_N \mathcal{O}^{medium}_{m}
  + C''' \mathcal{O}^{fast}_{k} + \ldots ,
  \label{comm3}
\end{equation}
where dots denote other fast operators which are already present in our system.
Equations of type \eqref{comm1}, \eqref{comm2} and \eqref{comm3} now form 
the closed system.
This system can be schematically written as
\begin{subequations}
  \begin{align}
  i \, \dot{x}_i &= \sum_j \left( \epsilon \, a_{i j} \, x_j + \epsilon \, b_{i j} \, y_j\right),
  \label{schematic_slow}\\
  i \, \dot{y}_k &= - E_k \,y_k+\sum_l \left(   \epsilon \, c_{k l} \, x_l    
  + \epsilon \, d_{k l} \, y_l\right), \label{schematic_fast}
  \end{align} \label{sys_schematic}\end{subequations}
where $x_i$ are the slow operators~\eqref{slow_comb} (their derivatives are proportional
to small parameter $\epsilon$), while variables $y_k$ are fast.
All coefficients $a, b, c, d$ are time dependent functions of order of unity
and $E_k$ are combinations of energies of HNLs and active neutrinos of type
$E_N+E_{\nu_\alpha}$, $E_{\nu_\alpha} - E_{\nu_\beta}$, etc.
Note that there is no summation over $k$ in~\eqref{schematic_fast}.
For the sake of clarity we have 
assumed the following power counting,
both Majorana mass differences and  Yukawas times the Higgs vacuum expectation value are
proportional to a small dimensional parameter 
$\Delta M , h\cdot \langle \Phi \rangle \propto \epsilon$.

We will show now, that the fast oscillations can be averaged or---in the language of effective 
theories---integrated out in the way that the final system describes the slow evolution of
operators.

Let us first consider the system at moment $\ov{t}$.
With  the fixed  $x_i(\ov{t}) = \ov{x}_i$ eqs~\eqref{schematic_fast} read
\begin{equation}
  i \, \dot{y}_k = - E_k \,y_k + \sum_l \left(    \epsilon \, c_{k l} \, \ov{x}_l  + 
  \epsilon \, d_{k l} \, y_l \right).
  \label{sys_fast}
\end{equation}
We choose a time interval $t \in [\ov{t}, \ov{t}+T]$,
such that $1/E\ll T\ll1/\epsilon$, where $E$ is of order of $E_N$ or $E_{\nu_\alpha}$.
The first inequality means that  $x_i(t)$ does not change significantly on this time interval.
We  solve the system~\eqref{sys_fast} on this  interval and denoted 
the solution as $\tilde{y}_k(t, \ov{x})$.
The second inequality allows us to exclude the fast oscillations from
$\tilde{y}_k(t, \ov{x})$ by means of the averaging
\begin{equation}
  \ov{y}_k(\ov{t}, \ov{x}) = \frac{1}{T}
  \int_{\ov{t}}^{\ov{t}+T} dt \, \tilde{y}_k(t,\ov{x}).
  \label{aver_sol}
\end{equation}
One can show that
for the system~\eqref{sys_schematic} originating from the commutation
of~\eqref{slow_comb} with the full Hamiltonian~\eqref{fullH}, averaging~\eqref{aver_sol} gives
\begin{equation}
  \ov{y}_k(\ov{t}, \ov{x}) = \sum_l \frac{ \epsilon \, c_{k l} \, \ov{x}_l}{E_k}.
  \label{aver_sol_expl}
\end{equation}

In practice, we first integrate out the interactions with medium. 
This allows us to express the operators $\mathcal{O}^{medium}_{m}$ in the
r.h.s. of \eqref{comm2} in terms of $\mathcal{O}^{fast}_{k}$. We are working in the leading order in small parameters $\epsilon$, as a result, it is only
the first term in the r.h.s. of \eqref{comm2} which should be modified. This modification can be summarized as
\begin{equation}
  \begin{aligned}
  E_\nu - E_N &\to E_\nu - E_N - \frac{1}{k}\bar{u}_h(k) \slashed{\Sigma}(k) u_h(k),\\
  E_\nu + E_N &\to E_\nu + E_N - \frac{1}{k}\bar{u}_h(-k) \slashed{\Sigma}(k) u_h(-k),
  \end{aligned}
  \label{E_shift}
\end{equation}
where $\Sigma(k)$ is the active neutrino self energy and we have suppressed the flavour indices of the active neutrinos.
The imaginary part of the self energy can be parametrized as~\cite{Weldon:1982bn} 
\begin{equation}
  \Im \slashed{\Sigma}(k) = \slashed{k} \, \Gamma_k/2  
  + \slashed{u} \, \Gamma_u/2 ,
  \label{self_energy}
\end{equation}
where $u^\mu = (1, 0,0,0)$ is the four-velocity of the medium.

Once the fast operators related to the medium are integrated out, we also get rid
of the fast operators of the type $\mathcal{O}^{fast}_{k}$. Eventually
one gets the closed set of equations in terms of the slow variables only.
These equations have the following generic form
\begin{equation}
i \, \dot{\ov{x}}_i(\ov{t}) = \sum_j  \epsilon \, a_{i j} \ov{x}_j +
\sum_{j,l}
\frac{ \epsilon^2 \,  b_{i j} c'_{j, l}}{E_\nu + E_N - i \left( \Gamma_u+ 2 k^0 \Gamma_k \right) } \, \ov{x}_j+
\sum_{j,l}
\frac{ \epsilon^2 \,  b_{i j} c''_{j, l}}{E_\nu - E_N - i \Gamma_u} \, \ov{x}_j,
  \label{sys_slow}
\end{equation}
where we have separated the coefficients $c_{j, l}$ into two groups, 
$c'$ and $c''$.
Eventually, the terms with $c''$ and $c'$ will lead to processes with and without fermion
number violation correspondingly, see eq.~\eqref{gamma_pm}.

\subsection{The final form of the equations} 
\label{sub:the_final_form_of_the_equations}

In order to obtain the system of kinetic equations in the matrix form we introduce
the convenient notations of ref.~\cite{Eijima:2017anv},
\begin{align}
&\begin{aligned}
\rho_{\nu_\alpha}&=\Tr[a^{\dagger}_{\nu_{\alpha}}(k) a_{\nu_{\alpha}}(k) \brho], \\
\rho_{\bar{\nu}_\alpha}&=\Tr[b^{\dagger}_{\nu_{\alpha}}(k) b_{\nu_{\alpha}} (k) \brho], 
\label{eq:notat_bnu} 
\end{aligned}\\
&\begin{aligned}
\rho_{N}&=\begin{pmatrix}
  \Tr[a^{\dagger}_{+}(k) a_{+}(k) \brho] & \Tr[a^{\dagger}_{+}(k) b_{-}(k)\brho] \\
  \Tr[b^{\dagger}_{-}(k) a_{+}(k) \brho] & \Tr[b^{\dagger}_{-}(k) b_{-}(k) \brho]
\end{pmatrix}, 
\label{eq:notat_N} 
\end{aligned}\\
&\begin{aligned}
\rho_{\bar{N}}&=\begin{pmatrix}
  \Tr[a^{\dagger}_{-}(k) a_{-}(k) \brho] & \Tr[a^{\dagger}_{-}(k) b_{+}(k) \brho] \\
  \Tr[b^{\dagger}_{+}(k) a_{-}(k) \brho] & \Tr[b^{\dagger}_{+}(k) b_{+}(k) \brho]
\end{pmatrix}.
\label{eq:notat_bN}
\end{aligned}
\end{align}
Using these notations we arrive at the following kinetic equations 
\begin{subequations}
\begin{eqnarray}
i \, \frac{d\rho_{\nu_\alpha}}{dt} 
&=& - i \, \Gamma_{\nu_\alpha} \rho_{\nu_\alpha}
    + i \, \Tr[\tilde{\Gamma}_{\nu_\alpha} \, \rho_{\bar{N}}], 
\label{eq:KEnu}\\
i \, \frac{d\rho_{\bar{\nu}_\alpha}}{dt} 
&=& - i \, \Gamma_{\nu_\alpha}^\ast \rho_{\bar{\nu}_\alpha}
    + i \, \Tr[\tilde{\Gamma}_{\nu_\alpha}^\ast \, \rho_{N}], 
\label{eq:KEnubar}\\
i \, \frac{d\rho_{N}}{dt} 
&=& [H_N, \rho_N]
    - \frac{i}{2} \, \{ \Gamma_{N} , \rho_{N}  \}
    + i \, \sum_\alpha \tilde{\Gamma}_{N}^\alpha \rho_{\bar{\nu}_\alpha},  
\label{eq:KEN}\\
i \, \frac{d\rho_{\bar{N}}}{dt} 
&=& [H_N^\ast, \rho_{\bar{N}}]
    - \frac{i}{2} \, \{ \Gamma_{N}^\ast , \rho_{\bar{N}} \}
    + i \, \sum_\alpha (\tilde{\Gamma}_{N}^\alpha)^\ast \rho_{\nu_\alpha}.
\label{eq:KENbar}
\end{eqnarray}
\label{KE_1}\end{subequations}
Expressions for the rates and the effective Hamiltonian are present in 
section~\ref{sec:production_of_the_baryon_asymmetry} so we do not repeat them here.

Notice that $\rho_N$, $\rho_{\bar{N}}$, $\rho_{\nu_\alpha}$ and 
$\rho_{\bar{\nu}_\alpha}$ in eqs.~\eqref{KE_1} depend on momentum
so there is fact a set of equations for each momentum mode.
During the whole period of the asymmetry generation, the leptons are in thermal 
equilibrium and different momentum modes
communicate to each other. Therefore, the appropriate variables for the r.h.s.
of equations~\eqref{KE_1} are the chemical potentials.
Subtracting~\eqref{eq:KEnubar} from~\eqref{eq:KEnu} and introducing the Fermi-Dirac
distribution function for massless neutrino $f_\nu = 1/\left( e^{k/T}+1 \right) $
we can rewrite (in the limit of small chemical potentials) 
equations~\eqref{KE_1} in the following form~\cite{Eijima:2017cxr}
\begin{subequations}
\begin{align}
i \frac{d n_{\Delta_\alpha}}{dt}
&= -  2 i \frac{\mu_\alpha}{T} \int \frac{d^{3}k}{(2 \pi)^{3}} \Gamma_{\nu_\alpha} f_{\nu} (1-f_{\nu})  \, 
    + i \int \frac{d^{3}k}{(2 \pi)^{3}} \left( \, \text{\text{Tr}}[\tilde{\Gamma}_{\nu_\alpha} \, \rho_{\bar{N}}]
    -  \, \text{\text{Tr}}[\tilde{\Gamma}_{\nu_\alpha}^\ast \, \rho_{N}] \right),\label{kin_eq_a0}
\\
i \, \frac{d\rho_{N}}{dt} 
&= [H_N, \rho_N]
    - \frac{i}{2} \, \{ \Gamma_{N} , \rho_{N} - \rho_N^{eq} \} 
    - \frac{i}{2} \, \sum_\alpha \tilde{\Gamma}_{N}^\alpha \, \left[ 2 \frac{\mu_\alpha}{T} f_{\nu} (1-f_{\nu}) \right],\label{kin_eq_b0}
\\
i \, \frac{d\rho_{\bar{N}}}{dt} 
&= [H_N^\ast, \rho_{\bar{N}}]
    - \frac{i}{2} \, \{ \Gamma_{N}^\ast , \rho_{\bar{N}} - \rho_N^{eq} \} 
    + \frac{i}{2} \, \sum_\alpha (\tilde{\Gamma}_{N}^\alpha)^\ast \, \left[ 2 \frac{\mu_\alpha}{T} f_{\nu} (1-f_{\nu}) \right],
\label{kin_eq_c0}
\end{align}
\label{KE_1a0}\end{subequations}
where $\rho_{N}^{eq}$ is the equilibrium distribution function of 
HNLs.\footnote{The equilibrium
distribution function appears as a result of application of the detailed balance principle.}
Note that  the r.h.s. of eq.~\eqref{kin_eq_a0} is written in terms of the density
of the $\Delta_\alpha=L_\alpha-B/3$, where $L_\alpha$ are the lepton numbers and
$B$ is the total baryon number. These combinations are not affected by the fast 
sphaleron processes and changes only due to interactions with HNLs, therefore 
their derivatives are equal to the derivatives of the lepton number densities $n_{L_\alpha}$.



 \section{Rates in the symmetric phase} 
 \label{sec:rates_in_the_symmetric_phase}
 In this section we describe how the rates entering eqs.~\eqref{KE_1a0}
 are defined. For completeness we also include the fermion number violating 
 Higgs decays and inverse decays in the symmetric phase.
 

\section{Benchmark points} 
\label{sec:benchmark_points}
In this chapter we present several parameter sets along with 
the corresponding values of the $Y_B$. These sets can be used to compare 
the numerical results among different groups. The full datasets could 
be downloaded from~\cite{IT}.
Note that we use the latest global fit to neutrino data~\cite{nufit}.

\begin{table}[htb!]
\begin{center}
  \begin{tabular}{| c | c | c | c | c | c |}
    \hline
   $M$, GeV & $\Delta M$, GeV  & $\imw$ & $\rew/\pi$ & $\delta/\pi$ &
    $\eta/\pi$ \\ \hline
5.0000e-01 & 5.9794e-09 & 5.3897e+00 & 6.4803e-01 & 1.8599e+00 & 3.5291e-01  \\
1.0000e+00 & 5.3782e-09 & 5.2607e+00 & 8.3214e-01 & 1.2708e+00 & 1.7938e+00  \\
2.0000e+00 & 3.0437e-09 & 5.5435e+00 & 1.6514e+00 & 1.6384e+00 & 7.5733e-01 \\
5.0000e+00 & 1.7945e-09 & 5.229e+00 & 1.7753e+00 & 1.4481e+00 & 1.2070e+00  \\ 
1.0000e+01 & 2.7660e-09 & 4.4442e+00 & 8.4146e-01 & 1.7963e+00 & 9.2261e-01\\
\hline
\end{tabular}
\end{center}
\caption{\label{NHbenchmarks} Parameter sets leading to the observed value of the BAU, NH case.}
\end{table}

\begin{table}[htb!]
\begin{center}
  \begin{tabular}{| c | c | c | c | c | c |}
    \hline
   $M$, GeV & $\Delta M$, GeV  & $\imw$ & $\rew/\pi$ & $\delta/\pi$ &
    $\eta/\pi$ \\ \hline
5.0000e-01 & 6.0739e-09 & 5.3788e+00 & 1.6652e+00 & 1.8721e+00 & 1.5305e+00 \\
1.0000e+00 & 8.3058e-09 & 4.9049e+00 & 9.1389e-02 & 1.5365e+00 & 4.7998e-01  \\
2.0000e+00 & 4.9537e-09 & 4.6975e+00 & 5.4153e-01 & 1.5263e+00 & 1.7442e+00 \\
5.0000e+00 & 2.3906e-09 & 4.1620e+00 & 2.7838e-01 & 1.2930e+00 & 9.9034e-01 \\ 
1.0000e+01 & 2.7097e-08 & -4.2412e-01 & 2.5429e-01 & 1.6901e+00 & 1.6839e-01 \\
\hline
\end{tabular}
\end{center}
\caption{\label{IHbenchmarks} Parameter sets leading to the observed value of the BAU, IH case.}
\end{table}


\bibliographystyle{JHEP}
\bibliography{BAUscanRefsUPDATED}

\end{document}